\documentclass[fleqn,usenatbib]{mnras}

\usepackage[T1]{fontenc}

\DeclareRobustCommand{\VAN}[3]{#2}
\let\VANthebibliography\thebibliography
\def\thebibliography{\DeclareRobustCommand{\VAN}[3]{##3}\VANthebibliography}

\usepackage{amsmath,amstext,physics}
\usepackage{float}
\usepackage[export]{adjustbox}
\usepackage{graphicx}
\usepackage{printlen,enumitem}
\usepackage{txfonts} 
\usepackage[figure,figure*]{hypcap} 
\usepackage{sidecap}
\usepackage{comment}
\usepackage{xcolor}
\usepackage[normalem]{ulem}

\newcommand\msun[1]{{\rm M}_{\odot}}
\newcommand{\Msun}{{\rm M}_{\odot}}

\newcommand{\Msunyr}{{\rm M}_{\odot}\,{\rm yr}^{-1}}
\newcommand{\kms}{{\rm km}\,{\rm s}^{-1}}

\title[FIRE winds fueling starbursts and quasars]{The role of galactic winds fueling central starbursts and quasars in the FIRE cosmological simulations}

\author[J. Mercedes-Feliz et al.]{Jonathan~Mercedes-Feliz,$^{1}$\thanks{E-mail: jonathan.mercedes\_feliz@uconn.edu}
Daniel Angl{\'e}s-Alc{\'a}zar,$^{1}$
Boon~Kiat Oh,$^{10,1}$
Rachel K. Cochrane,$^{2}$\newauthor
Sarah Wellons,$^{3}$
Alexander J. Richings,$^{4,5}$
Jorge Moreno,$^{6}$
Claude-Andr{\'e} Faucher-Gigu{\`e}re,$^{7}$\newauthor
Philip F. Hopkins,$^{8}$
and Du{\v s}an Kere{\v s}$^{9}$
\\
$^{1}$Department of Physics, University of Connecticut, 196 Auditorium Road, U-3046, Storrs, CT 06269-3046, USA\\
$^{2}$Jodrell Bank Centre for Astrophysics, University of Manchester, Oxford Road, Manchester M13 9PL, UK\\
$^{3}$Department of Astronomy, Van Vleck Observatory, Wesleyan University, 96 Foss Hill Drive, Middletown, CT 06459, USA\\
$^{4}$Centre for Data Science, Artificial Intelligence and Modelling, University of Hull, Cottingham Road, Hull, HU6 7RX, UK\\
$^{5}$E. A. Milne Centre for Astrophysics, University of Hull, Cottingham Road, Hull, HU6 7RX, UK\\
$^{6}$Department of Physics and Astronomy, Pomona College, 333 N. College Way, Claremont, CA 91711, USA\\
$^{7}$CIERA and Department of Physics and Astronomy, Northwestern University, 1800 Sherman Ave., Evanston, IL 60201, USA\\
$^{8}$TAPIR, Mailcode 350-17, California Institute of Technology, Pasadena, CA 91125, USA\\
$^{9}$Department of Physics, Center for Astrophysics and Space Sciences, University of California San Diego, 9500 Gilman Drive, La Jolla, CA 92093, USA\\
$^{10}$School of Physics, Korea Institute for Advanced Study, 85 Hoegiro, Dongdaemun-gu, Seoul 02455, Republic of Korea
}

\date{Accepted XXX. Received YYY; in original form ZZZ}

\pubyear{2025}

\begin{document}
\label{firstpage}
\pagerange{\pageref{firstpage}--\pageref{lastpage}}
\maketitle

\begin{abstract}
Central starbursts and Active Galactic Nuclei (AGN) are thought to be fueled by either galaxy interactions or secular processes in gravitationally unstable discs. We employ cosmological hydrodynamic simulations from the Feedback in Realistic Environments (FIRE) project to propose a new nuclear fueling scenario based on the transition that galaxies undergo from bursty to smooth star formation and from prominent global galactic winds to inefficient stellar feedback as they grow above $M_{\star}\sim 10^{10-10.5}\,\Msun$: the last major galactic wind event shuts down star formation, evacuates gas from the galaxy, and slows down gas accretion from the circumgalactic medium (CGM), creating a $\sim$$10^{10}\,\Msun$ pileup of gas in the inner CGM which later accretes coherently onto the galaxy, achieving a tenfold increase in inflow rate over pre-outflow conditions. We explicitly track the accumulation of gas along the outflow pathway owing to hydrodynamic interactions and show that $\sim$50\% of gas fueling the central $\sim$10--100\,pc over the subsequent $\sim$15\,Myr can be traced back to pileup gas having experienced $>$50\% change in infall velocity owing to the wind interaction. This galactic wind pileup effect may thus represent a significant fueling mode for compact starbursts and luminous AGN.
Galactic winds at earlier times or AGN-driven outflows can have qualitatively similar effects, but the pileup of gas driven by the last major galactic wind event refuels the galaxy precisely when the deepening stellar potential prevents further gas evacuation by stellar feedback, providing the ideal conditions for quasar fueling at the time when AGN feedback is most needed to regulate central star formation in massive galaxies at their peak of activity.
\end{abstract}




\section{Introduction} \label{sec:intro}
Galaxy-scale outflows driven by stellar feedback are a well-established feature in both cosmological simulations and observations of star-forming galaxies. Theoretical work using idealized simulations \citep[e.g.,][]{Smith2018,Kim2020,Schneider2020,Steinwandel2024} and high-resolution cosmological simulations such as those from the Feedback In Realistic Environments (FIRE)\footnote{\url{http://fire.northwestern.edu}} project \citep{Hopkins2014,Hopkins2018,Hopkins2023_fire3}, show that feedback from young stars --- including radiation pressure, stellar winds, photoionisation, and supernovae --- can drive bursty, multiphase outflows that significantly impact galaxy evolution \citep{Muratov2015,Agertz2016,Christensen2016,Muratov2017,Angles-Alcazar2017b,Sparre2017,Agertz2020,Pandya2021,Andersson2023}. These outflows are capable of ejecting a significant fraction of the interstellar medium (ISM), redistributing angular momentum, and enriching the circumgalactic medium \citep[CGM; ][]{Faucher-Giguere2023}. Observationally, a wide range of studies have detected such outflows via blueshifted absorption lines (e.g., Na\,I D, Mg\,II), extended emission (e.g., H$\alpha$, [O\,III]), and UV metal lines, confirming their prevalence across galaxy populations and cosmic time \citep[e.g.,][]{Martin2005, Rubin2014, Chisholm2017, Heckman2017,Roberts-Borsani2020}. These observations not only validate feedback implementations in simulations but also provide constraints on wind mass-loading factors, velocities, and spatial extent.

However, the fate of this ejected gas and its potential for recycling remains an open and crucial question. Several simulation-based studies address this by implementing particle tracking techniques. In previous works, tracking individual mass elements over time has proven to be a powerful tool to analyse galaxy formation simulations, investigating the origin of the stellar content and structure of galaxies \citep{Oser2010,Shipp2023,Gandhi2024}, CGM gas \citep{Hafen2019,Hafen2020}, gas accreted by black holes \citep[BHs;][]{Choi2024}, and the baryon cycle in galaxy evolution \citep{Christensen2016,Angles-Alcazar2017b,Grand2019,Ho2019,Tollet2019,Mitchell2020}. In the FIRE simulations, \citet{Angles-Alcazar2017b} use particle tracking to quantify the baryon cycle in haloes and demonstrated that a significant fraction of gas ejected by stellar feedback is later re-accreted, especially in more massive haloes. This ``wind recycling'' process contributes to prolonged star formation by allowing previously ejected gas to re-enter the galaxy disc on timescales ranging from $\sim$10\,Myr to $\sim$1\,Gyr. In fact, cosmological simulations often highlight wind recycling as a major contributor to the CGM gas reservoir and galaxy growth \citep[e.g.,][]{Oppenheimer2010,vandeVoort2017}. In principle, wind recycling could also have an effect on the amount and time dependence of gas fueling the central massive BH, but the plausible role of galactic winds triggering active galactic nuclei (AGN) remains an open question. 

The mass loading factor and velocity of winds, as well as their impact, depend on galaxy mass and redshift. In general, models favour higher mass-loading factors in low-mass galaxies, with outflow velocities increasing (though with weaker effect) in higher-mass systems. Importantly, high-resolution cosmological zoom-in simulations like FIRE predict that galaxies undergo a transition over time: low-mass galaxies at early times exhibit bursty star formation with relatively short bursts ($\sim$10\,Myr), whereas galaxies with $M_{\star} \gtrsim 10^{10}\,\Msun$ transition to steadier star formation and can even experience quenched periods of star formation lasting a few 100\,Myr due to stellar feedback-driven winds and disc settling \citep{Muratov2015,Muratov2017,Angles-Alcazar2017c,Sparre2017,Hayward2017,Faucher-Giguere2018,Pandya2021,Stern2021,Gurvich2023,Hopkins2023}. Bursty star formation is driven by shallow potential wells, where repeated cycles of gas accretion, star formation, and ejection are fuelled by clumpy, cold gas. Observationally, burst-driven deviations from the star-forming main sequence have been seen, particularly at high redshifts, where galaxies may still broadly follow the main sequence but move along it via intermittent bursts triggered by mergers or gas inflows \citep{Weisz2012,Rodriguez-Puebla2016,Tacchella2016,Emami2019,Faisst2019,Tacchella2020}. This transient behavior is mirrored in recent James Webb Space Telescope (JWST) results: for instance, \citet{Endsley2025} find that UV-faint galaxies at $z \sim 6$ frequently cycle through rapid starburst and downturn phases on $\sim$10--30\,Myr timescales—effectively creating `mini-quenched' states or brief departures from the main sequence, consistent with similar phenomena found in other recent works \citep{Looser2024,Looser2025,Cole2025,Covelo-Paz2025,Mintz2025,Simmonds2025}. Two main mechanisms have been proposed for the transition from bursty to steady star formation: (i) the deepening of the stellar gravitational potential well, and (ii) the virialisation of the CGM \citep{Muratov2015,Angles-Alcazar2017c,Ma2017a,Ma2017b,Stern2021,Yu2021,Hafen2022,Byrne2023,Gurvich2023}. These are not mutually exclusive, as previous studies have found the two mechanisms ``turn on'' nearly simultaneously and may be mutually reinforcing \citep[e.g.,][]{Byrne2023}.

Besides regulating galaxy growth, stellar feedback-driven winds can also strongly influence central BH growth. Simulations show that stellar feedback can suppress early BH accretion in low-mass galaxies \citep[e.g.,][]{Dubois2015,Angles-Alcazar2017c,Bower2017,Habouzit2017,Habouzit2021,Lapiner2021,Stern2021,Catmabacak2022,Byrne2023,Hopkins2023}. Observations in the local universe provide tentative support for a break in BH mass scaling relations, with undermassive BHs at $M_{\star}\lesssim 10^{10.5}\,\Msun$ \citep{Graham2013,Reines2015,Savorgnan2016,Davis2018,Davis2019,Sahu2019}, although the presence of this break can vary depending on galaxy type, property selection, and the intrinsic scatter in scaling relations \citep[e.g.,][]{Lasker2016,Nguyen2019,Schutte2019,Baldassare2020,Dullo2020}. Quasar luminosity functions also provide a hint for non-linear BH scaling relations \citep{Tillman2022}. Undermassive BHs can catch up once the galaxy grows, coinciding with the galaxy transitioning from bursty to steady star formation (described above, \citealt{Angles-Alcazar2017c,Catmabacak2022,Hopkins2023}). At that point, galactic wind material could become an additional source of fuel (rather than suppression) for BH growth via wind recycling.

Motivated by this, we propose that the last major stellar feedback–driven galactic outflow in the history of a massive galaxy can play a crucial role in fueling a central quasar at the time of this transition from bursty to steady star formation. We argue that recycling of gas from this event can fuel the nuclear region precisely when stellar feedback can no longer evacuate the gas reservoir, owing to the deepening stellar potential. Moreover, we suggest that the outflow’s expansion temporarily slows accretion onto the galaxy, creating a ``pileup’’ of gas in the inner CGM that can subsequently (re)accrete coherently, driving a larger gas inflow rate once the galactic gas disc reforms.

In this work, we use FIRE simulations to present a proof-of-concept analysis of this galactic wind pileup fueling scenario. We focus on a massive galaxy undergoing its last major global outflow event at $z \sim 2$, coincident with its transition from bursty to steady star formation, and employ a novel particle tracking technique to follow the expanding outflow and capture hydrodynamic interactions with the surrounding gas. We examine how this stellar feedback-driven outflow builds a large pileup of gas in the CGM and evaluate the subsequent inflow rate of pileup gas down to the nuclear region of the galaxy.

The outline of this paper is as follows: \S\ref{sec:methods} summarises the galaxy formation framework and our methodology to track the build up of gas impacted by the outflow; \S\ref{sec:overview} provides an overview of the simulations; \S\ref{sec:build_up} explores the buildup of gas; \S\ref{sec:flagged_properties} investigates the direct role of the outflow accumulating gas in the CGM and driving a coherent gas accretion event at later times; \S\ref{sec:discussion} discusses our results in the context of previous work; and \S\ref{sec:conclusions} provides a summary of our findings and the main conclusions of this work.

\begin{figure*}
\includegraphics[width = \textwidth]{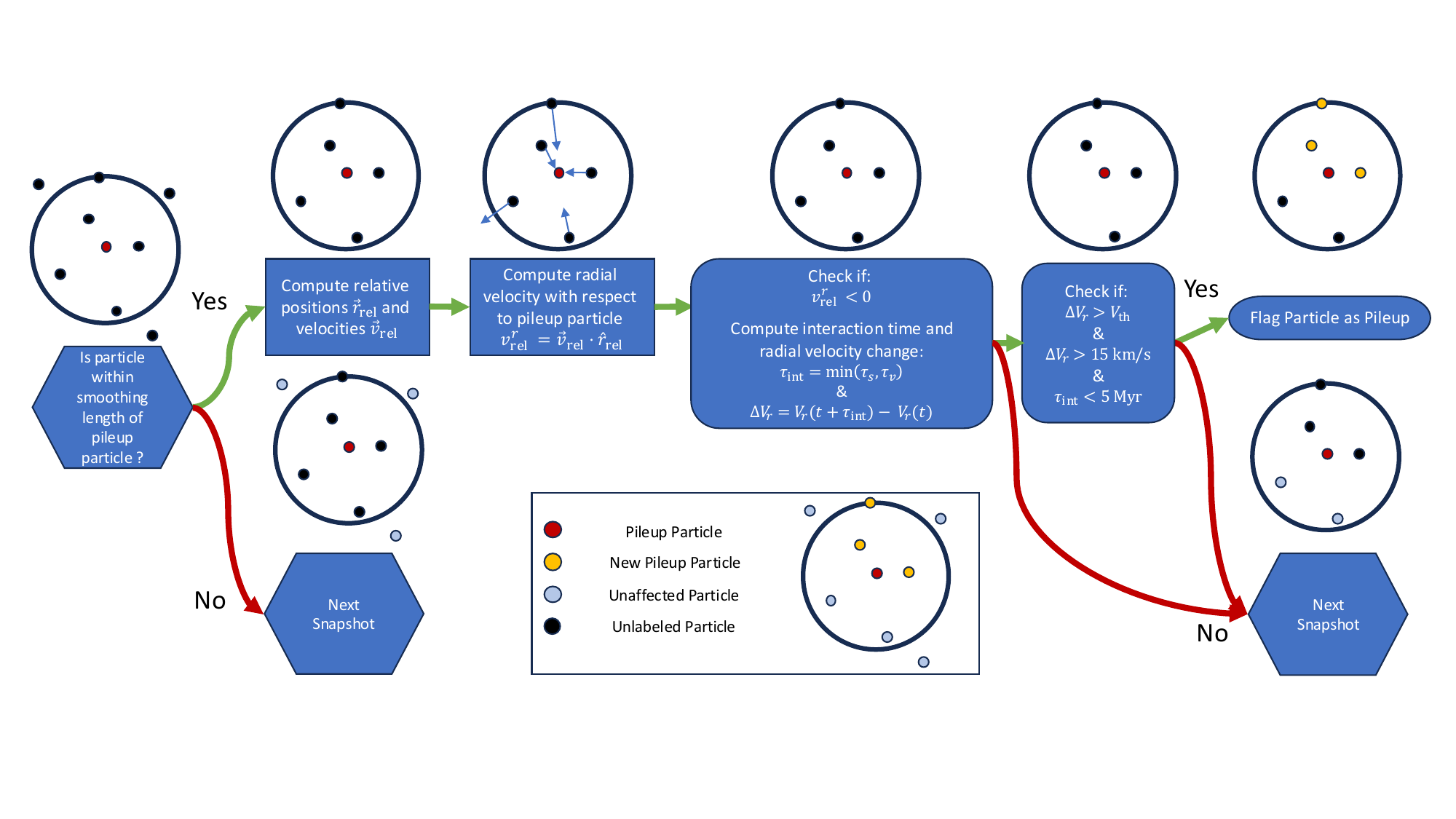}
\vspace*{-20mm}
\caption{Flowchart illustrating the PReS algorithm for identifying gas particles affected by the galactic wind pileup effect, including the decision-making criteria and the sequential steps involved.  The initial set of ``pileup particles'' is defined as all gas within the central 2\,kpc of the galaxy in the snapshot before the last global galactic outflow ($\Delta t \equiv -20\,{\rm Myr}$), corresponding to gas fully ejected by the time ($\Delta t \equiv 0\,{\rm Myr}$) at which the global galactic wind is first identified. Surrounding gas elements can be flagged in subsequent snapshots by interacting hydrodynamically with previously identified pileup particles. See text for details.}
\label{fig:flagging_schematic} 
\end{figure*}

\section{Methods}
\label{sec:methods}

\subsection{FIRE-2 galaxy formation model} \label{subsec:FIRE2model}

Our primary simulation is part of the FIRE project, specifically the ``FIRE-2'' galaxy formation physics implementation \citep{Hopkins2018}. The simulations use the $N$-body and hydrodynamics code GIZMO\footnote{\url{http://www.tapir.caltech.edu/~phopkins/Site/GIZMO.html}} in its ``meshless finite mass'' (MFM) hydrodynamics mode \citep{Hopkins2015gizmo}, a Lagrangian Godunov formulation which sets both hydrodynamic and gravitational (force-softening) spatial resolution in a fully-adaptive Lagrangian manner, with fixed mass resolution. As outlined in \citet{Hopkins2018}, we include cooling and heating from $T=10-10^{10}\,{\rm K}$; star formation in locally self-gravitating, dense ($n_{\rm H}\geq n_{\rm H, th} \equiv 1000\,{\rm cm}^{-3}$), molecular, and Jeans-unstable gas; and stellar feedback from OB \&\ AGB mass-loss, Type Ia \&\ II Supernovae (SNe), and multi-wavelength photo-heating and radiation pressure; with each star particle representing a single stellar population with known mass, age, and metallicity with all stellar feedback quantities and their time dependence directly taken from the \textsc{starburst99} population synthesis model \citep{Leitherer1999}.

\subsection{Massive star-forming galaxy at cosmic noon} \label{subsec:initialconditions}
We primarily focus on the massive FIRE-2 halo \textbf{A4} from \citet{Angles-Alcazar2017c}, reaching a mass $M_{\rm halo}\sim 10^{12.5}\,{\rm M}_{\odot}$ at $z=2$ and initially evolved down to $z=1$ including on-the-fly BH growth driven by gravitational torques \citep{Hopkins&Quataert2011,Angles-Alcazar2013,Angles-Alcazar2015,Angles-Alcazar2017a} but without AGN feedback. 
The central massive BH is modelled as a collisionless particle with an initial mass of $M_{\rm BH}=10^{8}\,\Msun$, located at the centre of the main simulated galaxy. Given that the BH mass is much larger than both the baryonic and dark matter particle masses, the BH dynamics within the galaxy are fully resolved without the need for artificially forcing the BH to maintain its central position \citep[e.g.,][]{Bahe2022}. As the BH grows through gravitational-torque accretion, mass conservation is ensured through stochastic swallowing of gas particles within the interaction kernel of the BH \citep[defined to contain $\sim$256 particles;][]{Angles-Alcazar2017c}.

This halo was first simulated with the FIRE-1 model as part of the MassiveFIRE suite \citep{Feldmann2016,Feldmann2017}. Here we use a resimulation of the FIRE-2 version of halo \textbf{A4}, which does not include AGN feedback, saving data snapshots at intervals of $\delta t\approx 0.2\,{\rm Myr}$ for $\sim$100\,Myr of total evolution time around $z\sim 2.28$. Other than this, the simulated halo adopts the same baryonic (gas and stellar) mass resolution $m_{\rm b}=3.3\times 10^{4}\,{\rm M}_{\odot}$ and dark matter mass resolution $m_{\rm DM}=1.7\times 10^{5}\,\Msun$, as well as gravitational force softenings $\epsilon_{\rm gas}^{\rm min}=0.7\,{\rm pc}$, $\epsilon_{\star}=7\,{\rm pc}$ and $\epsilon_{\rm DM}=57\,{\rm pc}$ for the gas (minimum adaptive force softening), stellar, and dark matter components. We assume a $\Lambda$CDM cosmology with parameters $H_{0}=69.7\,{\rm km}\,{\rm s}^{-1}\,{\rm Mpc}^{-1}$, $\Omega_{\rm M}=1-\Omega_{\Lambda}=0.2821$, $\Omega_{\rm b}=0.0461$, $\sigma_{8}=0.817$, and $n_{\rm s}=0.9646$ \citep{Hinshaw2013}.
In this work, we focus on halo \textbf{A4} in particular because of the availability of high-cadence snapshots, which allow us to track the detailed time evolution of inflows and outflows with a temporal resolution of $\delta t \approx 0.2$\,Myr. This makes it possible to follow the interaction between stellar-driven winds and the surrounding CGM with minimal ambiguity. While this high time-resolution motivates our choice of halo here, we note that the global star formation–driven outflows we analyze are typical in our broader suite of simulations, and we therefore expect the physical picture developed in this paper to apply more generally to similar systems.

In \citet{Mercedes-Feliz2023}, we selected the snapshot at $z=2.28$ to perform re-simulations including AGN winds \citep[see also][]{Cochrane2023a,Mercedes-Feliz2024}. At this time, the galaxy is experiencing a strong starburst phase, leading to the formation of an overcompact and overdense stellar component, as stellar feedback becomes insufficient and is no longer able to regulate ongoing star formation \citep[][]{Wellons2020,Parsotan2021}. Under these conditions, strong AGN winds can suppress star formation provided that the central BH can accrete efficiently \citep[][Angl{\'e}s-Alc{\'a}zar et al., in prep.]{Cochrane2023a,Mercedes-Feliz2023}. 
A separate set of Lagrangian hyper-refinement simulations of this galaxy have explicitly shown that strong gravitational torques from stellar non-axisymmetries can drive significant gas inflow rates down to sub-pc scales, explicitly powering a luminous quasar phase at this critical epoch \citep{Angles-Alcazar2021}. 

In this paper, we delve further back into this galaxy's history, examining the transition from bursty to steady star formation. In particular, we explore in detail the evolution and impact of the last major galactic outflow event powered by star formation, leading into the luminous quasar phase at $z\sim 2.28$. Here we define \( \Delta t = 0 \, \text{Myr} \) at the time when this galactic outflow has expanded to $\sim$1\,kpc (corresponding to $z\sim 2.32$),  with \( \Delta t \) representing the elapsed time since then.

\subsection{Particle tracking algorithm}
\label{subsec:track_methods}
To investigate the full impact of the galactic-scale outflow, we aim to identify gas with kinematic properties directly or indirectly affected by the outflow. This requires us to follow the initial outflowing material and flag any subsequent interactions with surrounding gas particles as it evolves in the simulation. Every particle in the simulation has a unique set of identifiers (IDs) that remain the same across snapshots, a feature that is especially useful for tracking particles across time.

Particle tracking has been used in previous works as a powerful tool to analyse galaxy formation simulations, investigating the origin of the stellar content and structure of galaxies \citep{Oser2010,Shipp2023,Gandhi2024}, CGM gas \citep{Hafen2019,Hafen2020}, gas accreted by BHs \citep{Choi2024}, and the baryon cycle in galaxy evolution \citep{Christensen2016,Angles-Alcazar2017b,Grand2019,Ho2019,Tollet2019,Borrow2020,Mitchell2020,Gebhardt2024}. In previous work, we use particle tracking to investigate the impact of AGN winds at the peak of star formation activity in halo \textbf{A4}, identifying local positive triggering of star formation coexisting with predominantly negative effects \citep{Mercedes-Feliz2023}, and tracing the formation of ultra-dense stellar clumps forming under the presence of strong AGN feedback \citep{Mercedes-Feliz2024}. Here we significantly extend the particle tracking method previously used in \citet{Mercedes-Feliz2023,Mercedes-Feliz2024} to quantify the plausible role of galactic-scale outflows as a new fueling mechanism to trigger a luminous quasar phase in galaxies.

To systematically track the evolution of outflowing particles and their interaction with the surrounding gas over time, we introduce and deploy the PReS (Propel-Reversal-Slowdown) algorithm. PReS iteratively identifies gas particles whose velocities are significantly altered by hydrodynamic interactions with the outflow. PReS identifies gas that is either propelled in galactic winds, reversed from inflowing to outflowing by gas dragging on CGM scales, or slowed down while accreting from the CGM. We quantify the ``pileup effect'' by applying PReS across simulation snapshots for $\sim$100\,Myr of evolution, where outflowing gas dynamically influences and entrains the surrounding material creating a pileup (or accumulation) of gas in the CGM that can later accrete onto the galaxy. To achieve this, we first identify the initial list of gas particles that originated the outflow. Since the outflow has already expanded to $\sim$1\,kpc by $\Delta t=0\,{\rm Myr}$, we examine the snapshot available immediately before the outflow onset ($\Delta t\sim -20\,{\rm Myr}$) when the galaxy was still intact. We define the {\it initial galactic outflow} as all gas particles located in the central 2\,kpc at $\Delta t\sim -20\,{\rm Myr}$ and not turned into stars within the next 20\,Myr, since tracking them forward in time can verify that in fact all of them have been ejected from the galaxy by $\Delta t=0\,{\rm Myr}$. This definition enables us to capture the faster component of the outflow, including gas particles that have already reached several kpc by $\Delta t = 0\,{\rm Myr}$, but our main results are insensitive to the details of the starting particle list. 

It is important to note that our initial selection of outflow particles does not rely on a velocity cut. Instead, we intentionally select all gas within the central 2\,kpc at the snapshot before the outflow. This comprehensive approach ensures that we include not only the fast-moving, expelled gas but also any slower-moving gas that is hydrodynamically influenced by the expanding outflow. By avoiding a velocity filter, we can later identify the full range of gas particles that have been impacted by the outflow, including those that may have remained within the ISM but experienced a significant change in their kinematics.
Initially defined outflow particles are tracked over time, along with their hydrodynamic interactions with surrounding gas particles. These gas particles can be flagged as affected by the pre-existing outflow and added to the list of particles tracked at the next snapshot. 
Our simulation's output cadence is sufficiently high (every $\sim$0.2\,Myr) to capture the relevant dynamical timescales of these interactions. Given the characteristic velocities of the gas and the typical smoothing lengths, the timescale for a particle to enter and exit a neighboring particle's influence is generally longer than the snapshot spacing, minimizing the possibility of missing short-lived interactions.
Following this process iteratively, the PReS algorithm allows us to identify gas particles dragged by the initial outflow all the way to the CGM, follow their subsequent evolution, and quantify their contribution to refueling the galaxy at later times. 

Figure \ref{fig:flagging_schematic} shows a flowchart summarising the main algorithmic elements of PReS, where blue nodes represent decision points (e.g., neighbor search, dynamical conditions), red arrows trace the path for particles that do not meet a given criterion, and green arrows indicate satisfied conditions towards flagging a given particle as newly affected by the outflow in the current snapshot. Hereafter, we refer to all particles affected by the outflow (and thus flagged and added to the growing list tracked in subsequent snapshots) as {\it pileup particles}. The PReS algorithm consists of the following key steps:

\begin{figure}
\includegraphics[width = \columnwidth]{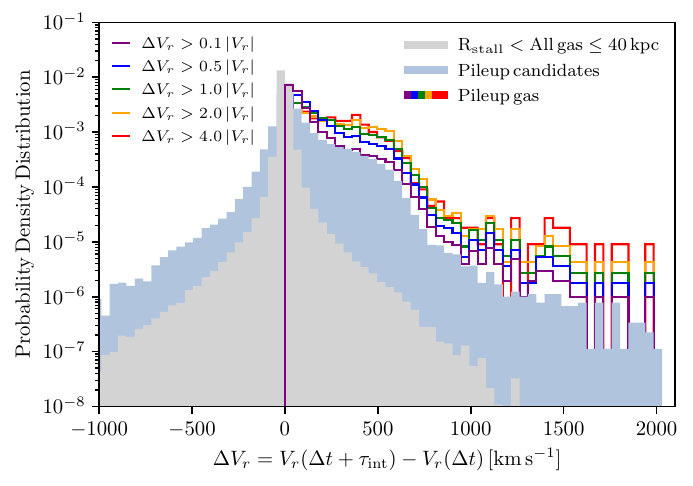}
\vspace*{-5mm}
\caption{Normalized probability distribution of radial velocity change measured at $\Delta t\sim 0\,{\rm Myr}$ for (1) all unaffected gas within $R_{\rm stall} < R \leq 40\,{\rm kpc}$ (grey), (2) pre-selected gas identified inside of the smoothing kernel of pileup gas (light blue), and (3) newly flagged pileup gas corresponding to different velocity thresholds (coloured lines). For pre-selected and pileup gas, velocity changes are calculated over the corresponding interaction time $\tau_{\rm int}$ for each particle ($0.2\,{\rm Myr} \lesssim \tau_{\rm int} \lesssim 5\,{\rm Myr}$). For comparison, velocity changes of unaffected gas are computed using $\tau_{\rm int} = 2\,{\rm Myr}$. Particles near pileup gas (light blue) exhibit larger changes in radial velocity on average compared to the background gas distribution (grey). Additionally, flagged pileup gas shows larger positive radial velocity changes with increasing velocity threshold, with average changes ranging from $\sim$$100\,{\rm km}\,{\rm s}^{-1}$ ($\beta =0.1$) to $\sim$$300\,{\rm km}\,{\rm s}^{-1}$ ($\beta=4$) and maximum velocity changes reaching $\sim$$2000\,\kms$. }
\label{fig:Vch_distribution} 
\end{figure}

\begin{figure*}
\includegraphics[width = \textwidth]{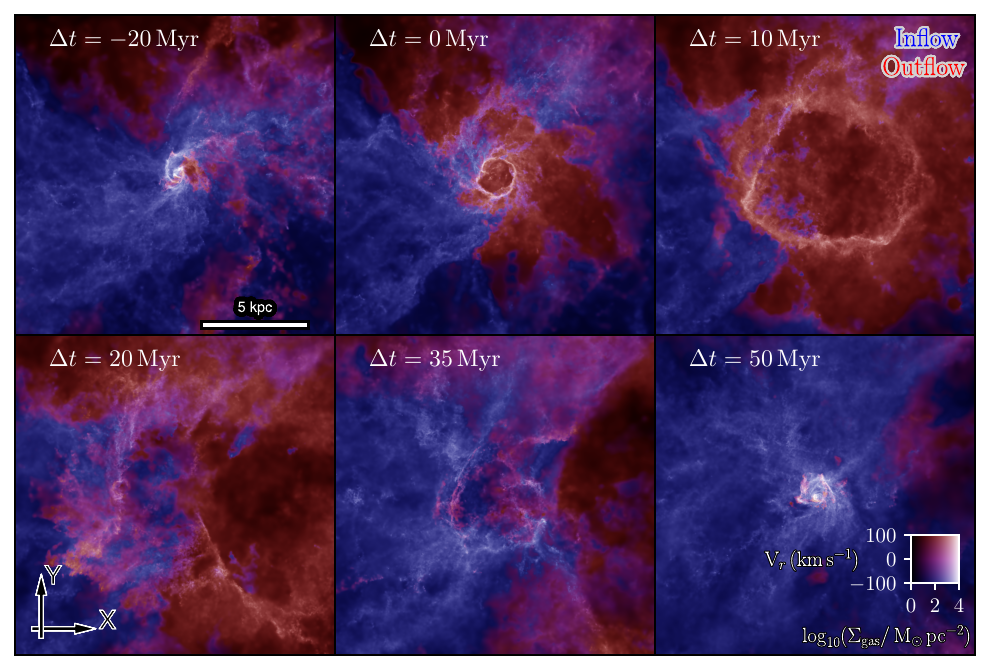}
\vspace*{-5mm}
\caption{Projected face-on gas mass surface density distribution and kinematics of gas in a 16\,kpc-wide region around a simulated massive, star-forming galaxy ($M_{\rm star} \sim 10^{10.5}\,\Msun$, ${\rm SFR} \sim 300\,\Msunyr$) at $z \sim 2.28$ over $\sim$70\,Myr. Inflowing gas is represented in blue hues, outflowing gas is represented in red hues, and the colour value and saturation correspond to the gas surface density logarithmically scaled.  
The last major galactic-scale outflow driven by stellar feedback fully evacuates the ISM within 1\,kpc at $\Delta t\equiv 0$\,Myr, expanding to larger scales while interacting with inflowing gas in the inner CGM (see larger cavity at $\Delta t=10$\,Myr). Once the outflow stalls (at $\Delta t\sim 20$\,Myr), the inflow component eventually dominates, and the deepening stellar potential enables the formation of a compact, high surface density gas disc at $\Delta t \sim 50$\,Myr which can no longer be disrupted by stellar feedback owing to gravitational confinement of winds.}
\label{fig:overview_panels} 
\end{figure*}

\begin{figure*}
\includegraphics[width = \textwidth]{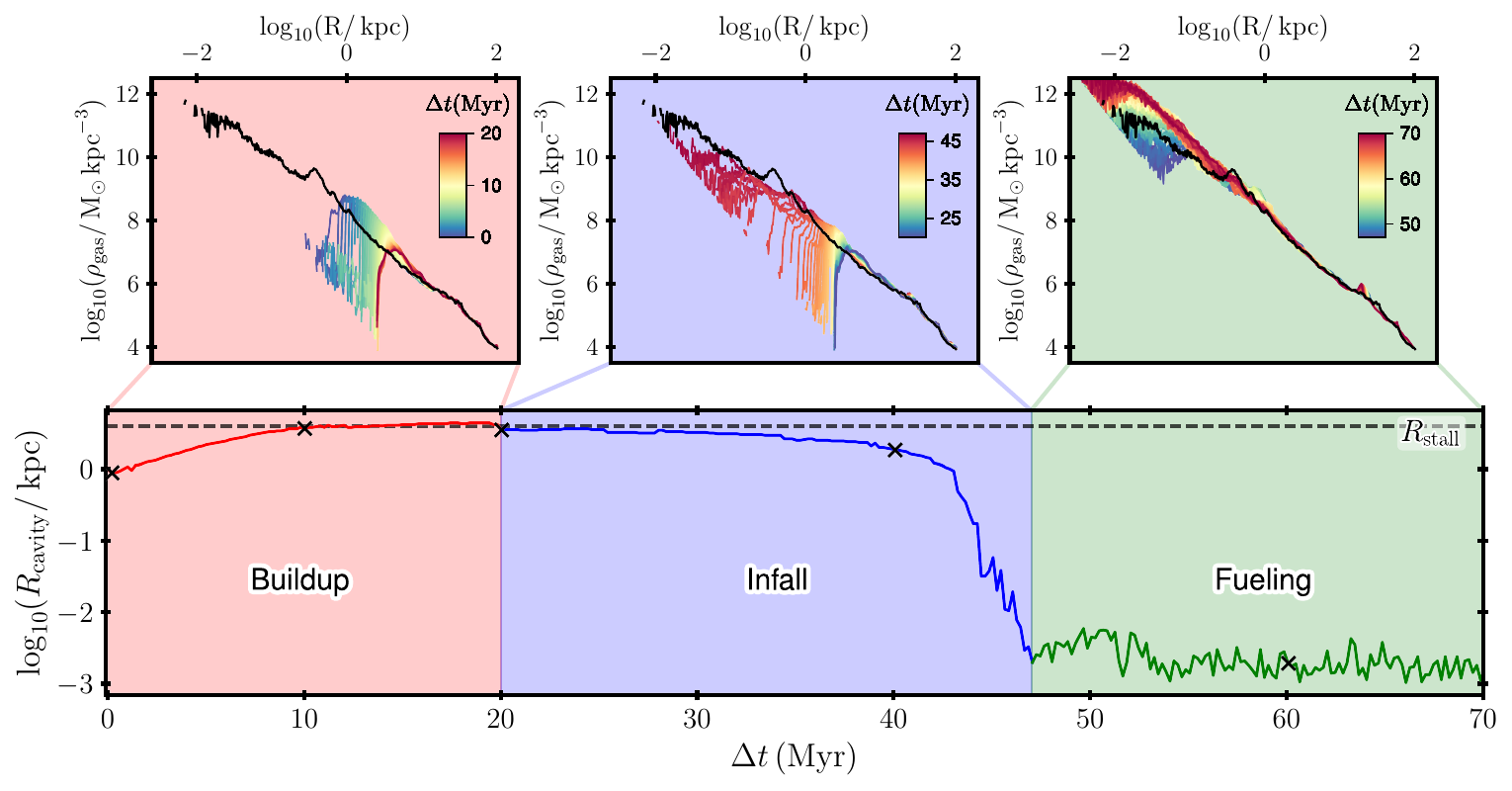}
\vspace*{-5mm}
\caption{Top: Radial profiles of the gas density, where the colour scale indicates time evolution within each phase and the black line is the radial density profile before the outflow event (at $\Delta t = -20$\,Myr; reproduced in the three panels for reference).
Bottom: Cavity radius as a function of time, following the expansion of the kpc-scale outflow up to the stalling radius $R_{\rm stall}\sim 4$\,kpc (``{\it Buildup}'' phase; red), the subsequent inflow of gas and reformation of the gas disc in the galaxy (``{\it Infall}'' phase; blue), and the ultimate fueling of the nuclear region (``{\it Fueling}'' phase; green).
The dashed horizontal line denotes $R_{\rm stall}$.
The x markers denote the times that are analysed in more detail in Figures~\ref{fig:flagged_maps},~\ref{fig:decomposed_densityradprofiles}, and \ref{fig:decomposed_flowradprofiles}. The nuclear gas density on $<$100\,pc scales during the {\it Fueling} phase greatly exceeds that of pre-outflow conditions.}
\label{fig:cavity_vs_time} 
\end{figure*}

\begin{enumerate}
    \item \textbf{Locate previously flagged particles:}\
    At snapshot $i$ (time $t$), we begin by identifying particles flagged as pileup in snapshot $i-1$ (time $t-\delta t$), including both previously found pileup particles and newly affected particles from the prior iteration.

    \item \textbf{Neighbor search within smoothing length:}\
    For each pileup particle, we search for gas particles within its smoothing length and consider them as potentially affected by the outflow. We denote the relative position between the pileup particle and a neighboring candidate in snapshot $i+1$ as $ \mathbf{r}_{\mathrm{rel}}$, and define their relative velocity as $\mathbf{v}_{\mathrm{rel}}$. We then compute the relative radial velocity, $v_{\rm rel}^{r}$, as the projection of relative velocity vector $\mathbf{v}_{\mathrm{rel}}$ onto the direction of $\mathbf{r}_{\mathrm{rel}}$, i.e., $v_{\mathrm{rel}}^{r} = \mathbf{v}_{\mathrm{rel}} \cdot \mathbf{r}_{\mathrm{rel}} / |\mathbf{r}_{\mathrm{rel}}|$.

    \item \textbf{Assess convergence and compute interaction time:}\
    We further select pairs of flagged and neighbor particles that are moving toward each other ($v_{\rm rel}^{r} < 0$) as a necessary condition to undergo an interaction. For each such pair, we compute two interaction timescales:
    \begin{align}
        \tau_v &= \frac{\|\Delta \mathbf{r}_{\mathrm{rel}}\|}{\lvert v_{\rm rel}^{r} \rvert}, \label{eq:tau_v}\\
        \tau_s &= \frac{\|\Delta \mathbf{r}_{\mathrm{rel}}\|}{\max(c_{s,\,\mathrm{pileup}}\,, c_s)},
        \label{eq:tau_s}
    \end{align}
    where $\tau_v$ is the time required for the particles to traverse their current separation at the relative radial velocity, and $\tau_s$ is the time for pressure perturbations to propagate across that separation. Here, $c_{s,\,\mathrm{pileup}}$ and $c_s$ are the sound speeds of the pileup and neighboring particles, respectively. The effective interaction time is defined as $\tau_{\mathrm{int}} = \min(\tau_v, \tau_s)$.

    \item \textbf{Check interaction thresholds:}\
    If $\tau_{\mathrm{int}} > 5\,\mathrm{Myr}$, we consider the interaction less likely to have an immediate dynamical effect and conservatively exclude the particle from the list of those potentially affected by the outflow. Interactions lasting longer than this are considered to be part of the broader, long-term galactic evolution and are less likely to represent a direct, immediate dynamical effect of the outflow. By imposing this limit, we ensure that our analysis is focused on particles that are directly and promptly affected by the outflow event, thereby preventing contamination from other, long-term physical processes. 
    For interactions with $\tau_{\rm int} < 5\,{\rm Myr}$, we then ask: does the neighbor particle show a significant change in its radial velocity over the interaction timescale? To quantify this, we compute the radial velocity of the neighbor relative to the BH at the start of the interaction, $V_{r}(t)$, and again after a time $\tau_{\rm int}$, $V_{r}(t+\tau_{\rm int})$. The latter is obtained directly from the simulation using the snapshot closest to $t+\tau_{\rm int}$. The change in radial velocity is then defined as $\Delta V_{r} = V_{r}(t+\tau_{\rm int}) - V_{r}(t)$. We compare $\Delta V_{r}$ against a velocity threshold $V_{\rm th} = \beta \lvert V_{\rm r}(t) \rvert$, where $\beta$ parametrizes the level of impact required to flag a significant pileup effect. We explore several values of $\beta$, from 10\% change in radial velocity ($\beta = 0.1$) up to a factor of four increase in radial velocity ($\beta = 4$) due to the interaction with the outflow. If $\Delta V_{r} > V_{\mathrm{th}}$ and $\Delta V_{r} > 15\, {\rm km}\,{\rm s}^{-1}$ (exceeding the sound speed of gas at $T=10^{4}\,{\rm K}$ to conservatively ignore smaller velocity changes), we flag the neighbor particle as a new pileup particle.

    \item \textbf{Update the pileup particle list:}\
    Newly flagged pileup particles in snapshot $i$ are added to the tracked particle list from snapshot $i-1$. The updated list is then carried forward to the next snapshot, and the procedure repeats. As multiple pileup particles can flag the same particle within a single snapshot, we record only the first occurrence since the identity of the flagging particle does not influence subsequent evolution.
\end{enumerate}

This iterative process systematically captures gas particles undergoing significant velocity changes (as required by the choice of $\beta$) at the time of interacting with previously identified outflowing particles, whether directly in contact with the initial outflow or indirectly through secondary interactions as more gas is dragged by the expanding outflow. 
Figure \ref{fig:Vch_distribution} illustrates the particle selection process in the PReS algorithm to identify pileup gas hydrodynamically impacted by the outflow. The normalized probability distribution of radial velocity changes, measured at $\Delta t \sim 0\,{\rm Myr}$, is presented for three different population of gas particles. We define the stalling radius, $R_{\rm stall}$, as the maximum cavity size before it begins to contract, which in this case occurs at $t_{\rm stall}\sim 20$\,Myr. The grey filled distribution represents all unaffected ``background'' gas within $R_{\rm stall} < R \leq 40\,{\rm kpc}$. The light blue filled distribution shows pre-selected gas identified within the smoothing kernel of pileup gas, representing initial candidates for hydrodynamic interaction. Finally, the coloured lines correspond to newly flagged pileup gas, differentiated by various $\beta$ velocity thresholds. For pre-selected and pileup gas, velocity changes are calculated over their individual interaction times ($\tau_{\rm int}$), which range from $\sim$0.2--5\,Myr. For comparison, velocity changes for unaffected gas are computed using a representative interaction time $\tau_{\rm int} = 2\,{\rm Myr}$ for all particles.

A clear distinction in radial velocity changes is observed between the gas hydrodynamically in contact with the outflow and the background gas distribution, with the former exhibiting, on average, larger changes. This demonstrates the algorithm's ability to isolate gas undergoing significant dynamic interactions. Notably, while the grey distribution of background gas shows a roughly symmetric spread of radial velocity changes, extending similarly to positive and negative values, the light blue distribution of candidate pileup particles is preferentially skewed towards positive radial velocity changes, indicating a net outward acceleration or redirection. Particles that are further identified and flagged as pileup consistently demonstrate larger and always positive radial velocity changes, increasingly so with higher values of the threshold velocity parameter $\beta$. The average velocity change for flagged pileup gas ranges from $\sim$$100\,\kms$ for $\beta = 0.1$ to $\sim$$300\,\kms$ for $\beta = 4$. Furthermore, as $\beta$ increases, a larger number of pileup particles experience velocity changes reaching up to $\sim$$2000\,\kms$. This quantitative trend confirms that higher $\beta$ values identify gas experiencing more pronounced hydrodynamic acceleration, validating the PReS algorithm's sensitivity to the degree of hydrodynamic influence from the outflow.

As an additional constraint to minimize possible spurious interactions, we further impose a threshold in radial distance ($R_{\rm cut} = 4\,{\rm kpc}$) and refrain from flagging pileup particles inside this radius once the outflow cavity stalls ($t_{\rm stall} \sim 20$ Myr). At this point, the outflow-induced cavity begins to shrink, an inflow-dominated phase emerges, and the gas disc reforms (see Figure~\ref{fig:cavity_vs_time}). The chosen $R_{\rm cut} = 4\,{\rm kpc}$ approximates the stalling radius $R_{\rm stall}$ (the maximum cavity size before contraction begins), inside of which gas would not be expected to be affected by the outflow at $t>t_{\rm stall}$.

Overall, the PReS algorithm enables us to track both initially outflowing gas from the galaxy as well as material dynamically influenced by subsequent interactions all the way to the CGM. By iterating across 483 snapshots over $100\,{\rm Myr}$, we construct a more complete picture of the outflow’s full extent and its effects on gas over a broad range of scales. Crucially in this work, PReS also enables us to track the subsequent evolution of pileup gas (affected by the outflow) and its role (re)accreting onto the galaxy at later times, potentially fueling the nuclear region and triggering a compact starburst and/or a luminous quasar phase.

\begin{figure*}
\includegraphics[width = 0.86\textwidth]{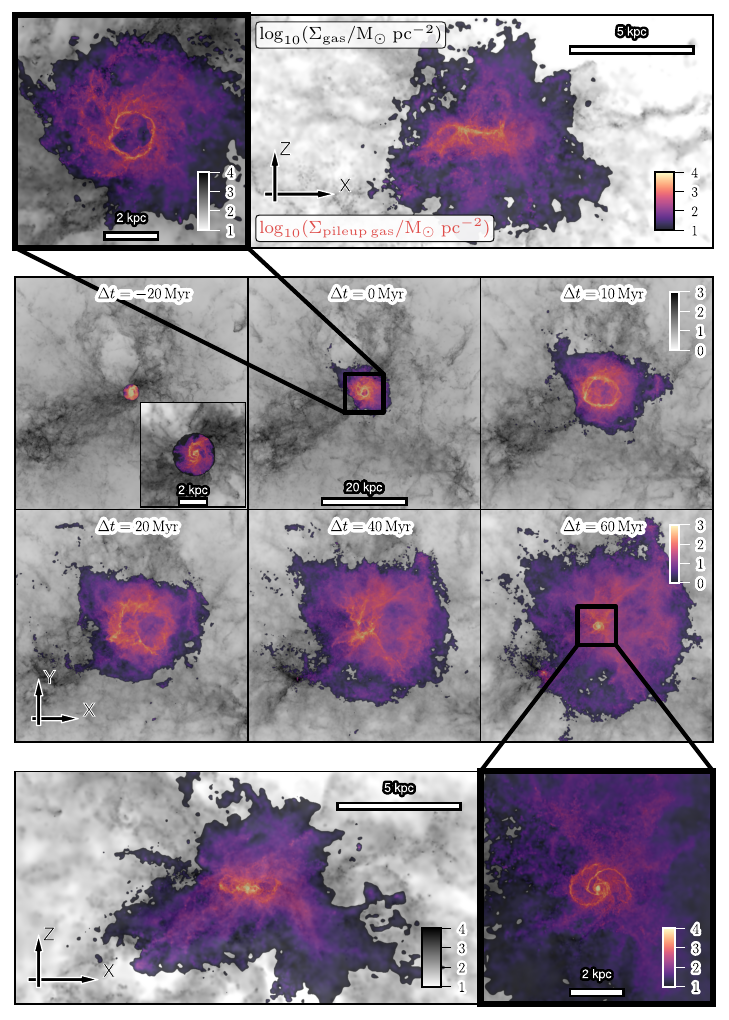}
\vspace*{-5mm}
\caption{Expansion of the global galactic wind and interaction with the infalling CGM, creating a pileup of gas that later accretes onto the galaxy and enables the formation of an ultra-dense nuclear gas disc. The middle panels show the projected gas mass surface density distribution (background grey scale) in the central 60\,kpc region over a 60\,Myr period since the onset of the last major galactic outflow. The propagation of galactic wind pileup gas is indicated by the purple-to-yellow colour scale, corresponding to gas particles flagged by the PReS algorithm as having been impacted by the pileup effect (in this case for $\beta=0.5$, i.e. experiencing $>$50\% change in radial velocity). The top and bottom rows show close-up views of the central 10\,kpc region at $\Delta t = 0$\,Myr and $\Delta t=60$\,Myr, respectively, in both the face-on and edge-on projections, showing that a substantial amount of gas hydrodynamically impacted by the wind contributes to the sub-kpc nuclear gas reservoir at later times.} 
\label{fig:flagged_maps} 
\end{figure*}

\begin{figure*}
\centerline{
\includegraphics[width = 0.33
\textwidth]{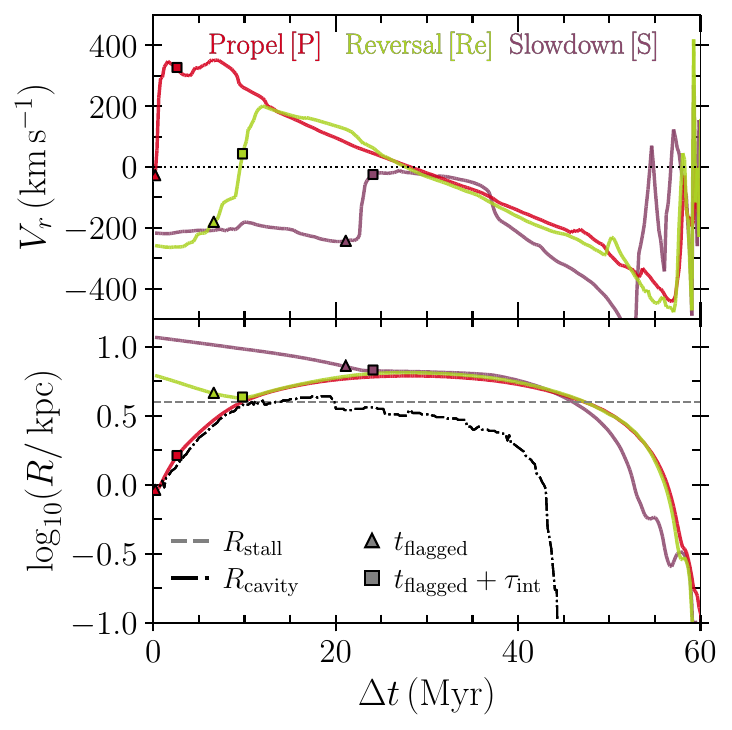}
\hspace{-2mm}
\includegraphics[width = 0.34\textwidth]{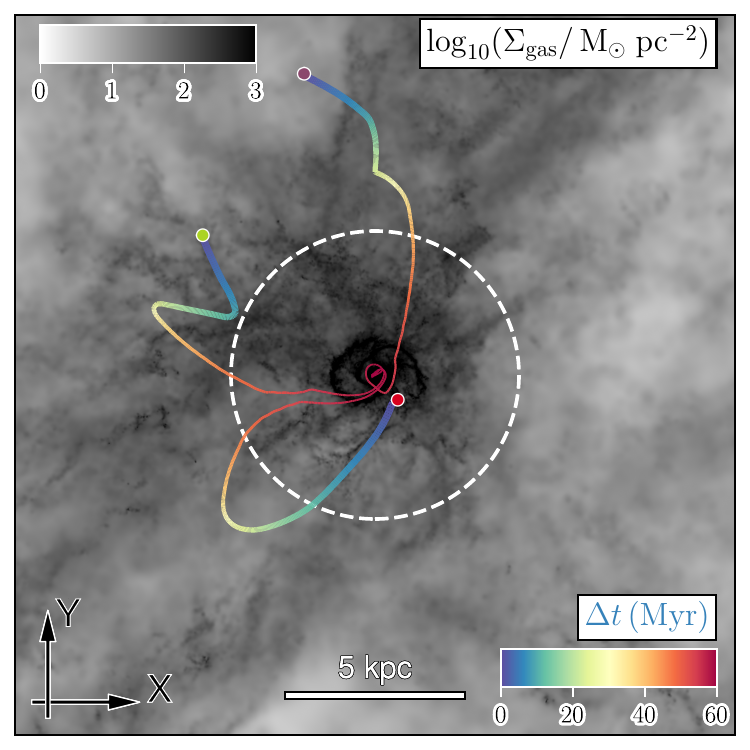}
\hspace{-2mm}
\includegraphics[width = 0.34\textwidth]{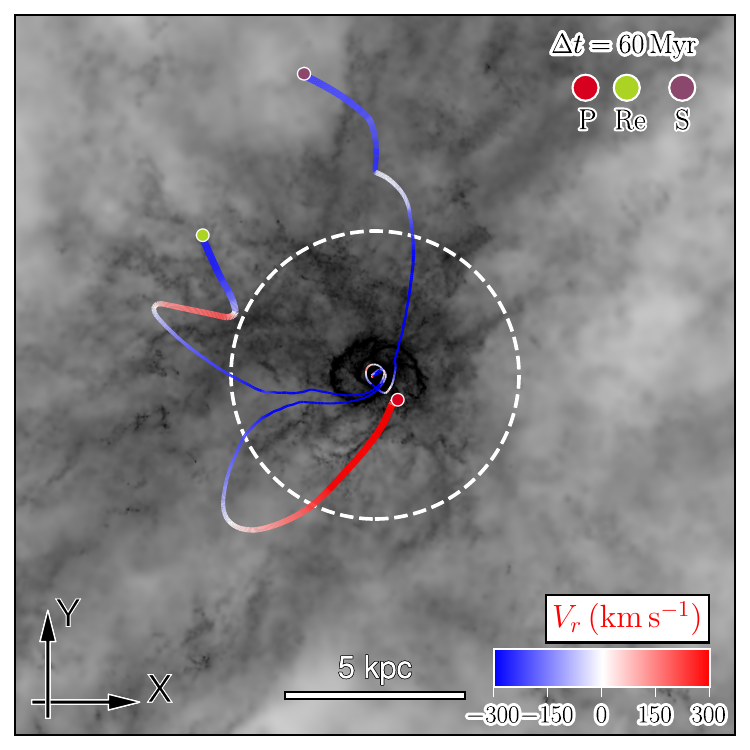} }
\caption{Left: Time evolution of radial velocity (top) and radial distance (bottom) for three representative gas particles that reach the central 100\,pc of the galaxy by $\Delta t \sim 60\,{\rm Myr}$. These particles illustrate three possible scenarios of pileup flagging in the PReS algorithm: Propel (P; red), Reversal (Re; green), and Slowdown (S; purple). The triangle markers indicate when the gas particles are flagged, while square markers denote the interaction time identified by the PReS algorithm. The dashed horizontal grey line represents the stalling radius, and the black dash-dotted line marks the cavity radius. Middle: Face-on projection of the gas mass surface density in the central (8\,kpc)$^{2}$ at $\Delta t\sim 60\,{\rm Myr}$. Trajectories for the same three particles are shown from $\Delta t = 0 \rightarrow 60\,{\rm Myr}$, with line colour indicating time evolution. Right: The same particle trajectories as in the middle panel, but with line colour representing radial velocity: inflowing gas in blue and outflowing gas in red. The white dashed circle marks the stalling radius. 
Identified pileup particles can reach the galaxy's central region through a variety of trajectories exhibiting different forms of impact by the pileup effect, including: (1) ISM gas propelled in galactic winds recycling back at later times, (2) the reversal of inflowing gas dragged by the outflow on CGM scales, and (3) the slow down of accreting gas from the CGM.  
}
\label{fig:PReS_trajectories} 
\end{figure*}

\section{Overview of inflow-outflow interaction} \label{sec:overview}
Figure~\ref{fig:overview_panels} presents images of the simulated galaxy, displaying face-on projected total gas mass surface density. The images are colour-coded by the mass-weighted radial velocity to distinguish between inflowing and outflowing gas components during $\sim$80\,Myr of evolution around the time of the last global galactic wind driven by stellar feedback for this galaxy.
This image utilises a hue-saturation-value (HSV) transformation where:
(i) Hue represents the radial velocity, encoding inflows and outflows across a spectrum from blue (inflow) to red (outflow);
(ii) Saturation is modulated by the gas mass surface density to enhance visibility of denser regions; and
(iii) Value is derived from the gas mass surface density to control brightness and contrast in the image.
This method effectively highlights the spatial distribution and dynamic state of gas influenced by the large-scale stellar feedback-driven outflow event. The radial velocity is defined relative to the centre of mass velocity of stars within the central 1\,kpc of the galaxy.

The $\Delta t=-20$\,Myr panel (top left) illustrates the pre-outflow state of the galaxy, in which the gas is characterised by a dense spiral morphology with the CGM dominated by an inflow component (noted by the extensive blue regions). At this stage, the galaxy maintains a steady gas inflow rate, providing material that later cools and fuels the central starburst driving the last large-scale stellar feedback-driven outflow in the simulated galaxy. In the subsequent panel ($\Delta t-0$\,Myr), we observe the onset of the outflow, which has at this point evacuated most of the gas from the central $\sim$1\,kpc region and disrupted the inflow of gas from the CGM, transitioning to an outflow-dominated state (predominantly red) for $\sim$20\,Myr. As the outflow expands, it is slowed down by the gas inflow from the CGM, leading to a dynamic interplay between the outflow and inflow components. This interaction deposits momentum from the outflow into the inflow, reducing their respective velocities. Ultimately, strong gas accretion on halo scales overcomes the outflow and the CGM becomes inflow dominated (predominantly blue) as the galaxy begins to reform at $\Delta t > 35$\,Myr.

The bottom right panel shows the newly reformed galaxy at $\Delta t=50$\,Myr, corresponding to the early stages of a subsequent starburst phase, where the galaxy evolves into a turbulent, clumpy, kpc-scale disc marked by intense star formation along fractured spiral arms and a dense nuclear region. Over time, continuous inflow of gas across scales replenishes the central region, transforming it into an ultra-compact, nuclear version. This phase reaches star formation rates (SFRs) of $\sim$400\,$\Msunyr$, but stellar feedback cannot regulate star formation nor further gas inflow onto the galaxy owing to the gravitational confinement of galactic winds by the deepening stellar potential.

Unlike previous cycles of blowout and accretion (due to the bursty nature of star formation), these extreme conditions lead to a sustained starburst phase without any further stellar feedback-driven outflows. Our simulation continues for an additional $\sim$100\,Myr, after its last major galactic outflow event, and during this time, stellar feedback remains ineffective at regulating the intense star formation and continued gas inflow. This inability of stellar feedback to halt the starburst and prevent the formation of an overly compact and dense stellar component highlights the necessity of another quenching mechanism such as AGN feedback \citep{Wellons2020,Parsotan2021}, which is expected to ultimately quench the galaxy and place it on the observed galaxy scaling relations at $z \sim 2$ \citep{Cochrane2023a,Mercedes-Feliz2023}.

\section{Gas Pileup driven by Galactic Outflow} \label{sec:build_up}

Figure~\ref{fig:cavity_vs_time} quantifies the impact of the outflow over a period of 70\,Myr in the radial gas density profile of the galaxy (top) and the corresponding size of the central cavity created by the outflow (bottom). We use spherical radial bins logarithmically spaced with width $\Delta\log_{10}(R) = 0.01\,{\rm dex}$ to compute the gas mass per unit volume as a function of distance to the centre of the galaxy. Based on these radial density profiles, we define the cavity size $R_{\rm cavity}$ as the distance from the centre at which the density profile reaches its maximum density at each time.

We define three separate phases given the time evolution of $R_{\rm cavity}$ seen in the bottom panel, which we indicate with different colors: (i) the {\it Buildup} phase (red) corresponds to the initial $\sim$20\,Myr when the outflow is propagating and the central cavity is growing; (ii) the {\it Infall} phase (blue) denotes the period of time after the outflow has transferred most of its momentum and the gravitational inflow of gas begins to dominate, eventually driving gas down to the galaxy and refueling the central cavity; and (iii) the {\it Fueling} phase (green) is the last stage where the galaxy has reformed and the gravitational potential well has become deep enough that stellar feedback is no longer efficient in shutting down star formation or in ejecting gas out of the galaxy. 
During the {\it Buildup} phase of $\sim$20\,Myr, the cavity expands but at a decreasing rate, reaching a maximum size of $\sim$4\,kpc. This expansion coincides with a peak outflow rate exceeding $1000\,\Msunyr$, after which the outflow stalls for roughly 1\,Myr. As the outflow pushes the ISM, it becomes increasingly asymmetric, with fast outflowing gas expanding more efficiently along paths of least resistance (also seen in Figure~\ref{fig:overview_panels}). Consequently, the cavity's shape becomes elongated, with its effective size more closely aligned with the longer axis of the cavity. As the outflow dissipates and the cavity begins to close, the rate of closure is initially slower compared to the expansion rate in first $\sim$10\,Myr, but the final $\sim$1\,kpc cavity refills very quickly once the galaxy reforms at $\Delta t \gtrsim 40$\,Myr. Once the galaxy has reformed, the cavity size as defined here no longer corresponds to an outflow and instead reflects the turbulent dynamics of gas in the nuclear region of the galaxy in its {\it Fueling} phase.

The top panels in Figure~\ref{fig:cavity_vs_time} show the radial distribution of gas as a function of time. For each panel, we focus on the radial profiles that correspond to each of the three phases outlined above. We show radial profiles at every 0.4\,Myr to highlight the relevant trends, where earlier times appear blue and turn red as we go further in time (as indicated by the colour scale). We also show the initial gas distribution from the snapshot before the outflow ($\Delta t = -20\,{\rm Myr}$) as the black line for comparison. As the outflow evacuates the central 1\,kpc region, shutting down star formation, gas is compressed along the shock front and accumulates gas at larger distances. Gas continues to accrete from the CGM but at a slower rate, depositing gas alongside the outflow cavity and creating a pileup of gas. This reservoir of gas continues to grow as the outflow propagates further and the cavity grows larger (see the clear ring structure at $\Delta t=10$\,Myr and $\Delta t=20$\,Myr in Figure~\ref{fig:flagged_maps}). During the {\it Buildup} phase, the radial range $3<R<10\,{\rm kpc}$ accumulates more than twice the gas mass originally contained within the central 1\,kpc before the outflow. Once the simulation reaches the stalling time ($t_{\rm stall} \sim 20$\,Myr), the galaxy transitions to be inflow dominated and the cavity stalls, starting the {\it Infall} phase of the galaxy. Once the galaxy has reformed ($\Delta t\sim 47$\,Myr), the gas density profile becomes similar to pre-outflow conditions in less than $\sim$5\,Myr. In the following $\sim$20\,Myr, gas continues to flow into the nuclear region with the gas density reaching $>10^{11}\,\Msun\,{\rm kpc}^{-3}$ within 100\,pc, greatly exceeding the pre-outflow conditions.

\begin{figure*}
\includegraphics[width = \textwidth]{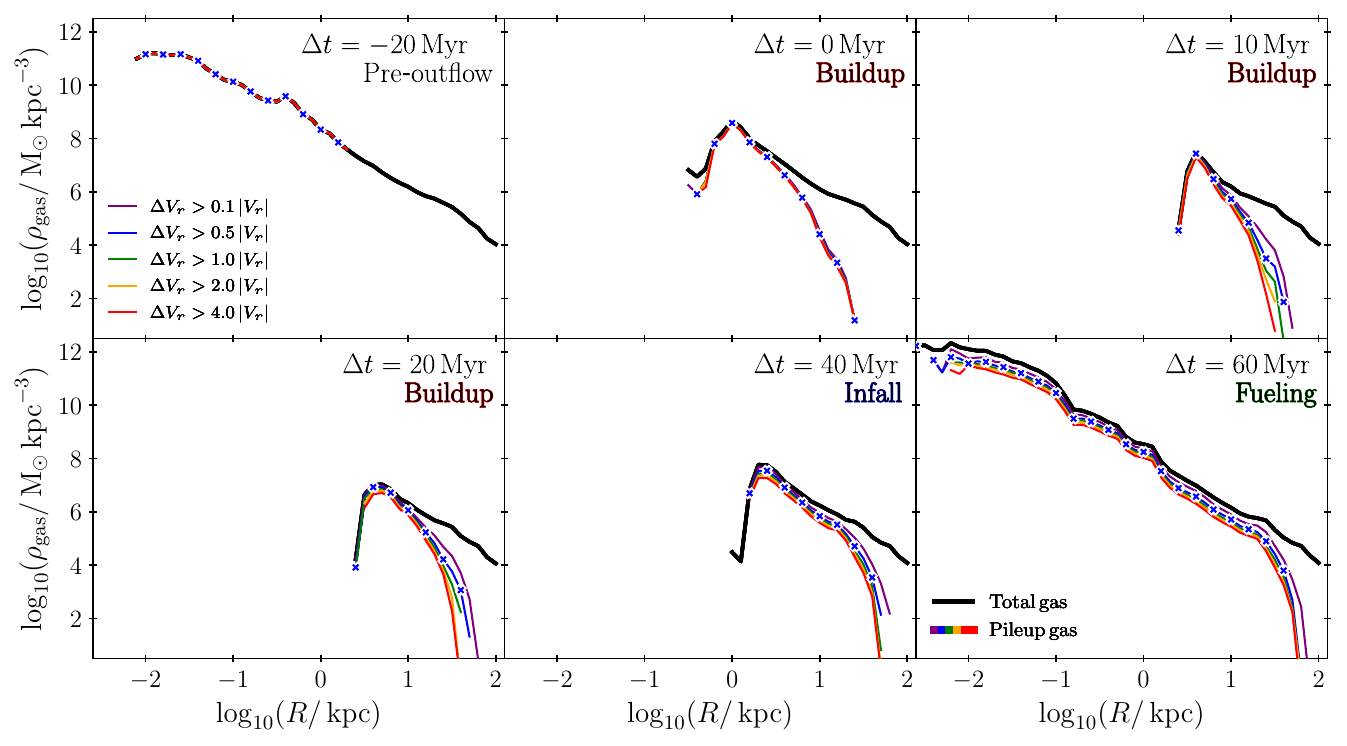}
\vspace*{-2mm}
\caption{Radial profile of the gas mass density ($\rho_{\rm gas}$) as it evolves from $\Delta t = -20 \rightarrow 60$\,Myr, corresponding to the same snapshots as were used to ilustrate the 2D projected gas distributions in Figure~\ref{fig:flagged_maps}. 
Black lines show the total gas profiles and the colour lines are for pileup gas only, showing the amount of gas impacted by the galactic wind pileup effect. Different colours indicate varying thresholds to measure the impact of the pileup effect, from requiring a minimum of 10\% change in radial velocity ($\Delta V_{r}>0.1\,|V_{r}|$; purple) up to more than a factor of four change in radial velocity owing to hydrodynamic interactions with the expanding galactic wind ($\Delta V_{r} > 4\,|V_{r}|$; red).
The x markers denote the $\beta =0.5$ threshold (blue), which we use as representative case in this work.
Pileup gas impacted by the wind dominates the inner few kpc during the {\it Buildup} ($\Delta t \lesssim 20\,{\rm Myr}$) and {\it Infall} ($\Delta t \sim 40\,{\rm Myr}$) phases, contributing a substantial fraction of gas across the full range $R\sim 10\,{\rm pc}-10\,{\rm kpc}$ at later times.}
\label{fig:decomposed_densityradprofiles} 
\end{figure*}

Using the PReS particle tracking algorithm described in \S\ref{subsec:track_methods}, we can further study the impact of the outflow creating a pileup of gas in the CGM and its eventual contribution to fueling nuclear star formation and BH accretion. We illustrate the propagation of gas that has been affected by the outflow in Figure~\ref{fig:flagged_maps}, showing face-on gas mass surface density projections for the central (60\,kpc)$^{3}$ region at different times (marked in Figure~\ref{fig:cavity_vs_time} for reference). The background grey colour scale shows the total gas mass distribution while the plasma colour scale overlaid indicates pileup gas, i.e., gas that has been flagged as impacted by the pileup effect. While we explore different velocity thresholds ($\beta$ values) to identify pileup gas, this figure specifically illustrates the propagation of the pileup effect for the $\beta=0.5$ case, where particles are flagged as affected by the outflow if they experience a radial velocity change of at least 50\% of their current radial velocity at the time of interacting with the outflow. We also conduct a detailed analysis of two snapshots, $\Delta t=0$ and $\Delta t=60$\,Myr, where we zoom into the central 10\,kpc region for a closer view of the central galaxy in face-on and edge-on orientations.

\begin{figure*}
\includegraphics[width = \textwidth]{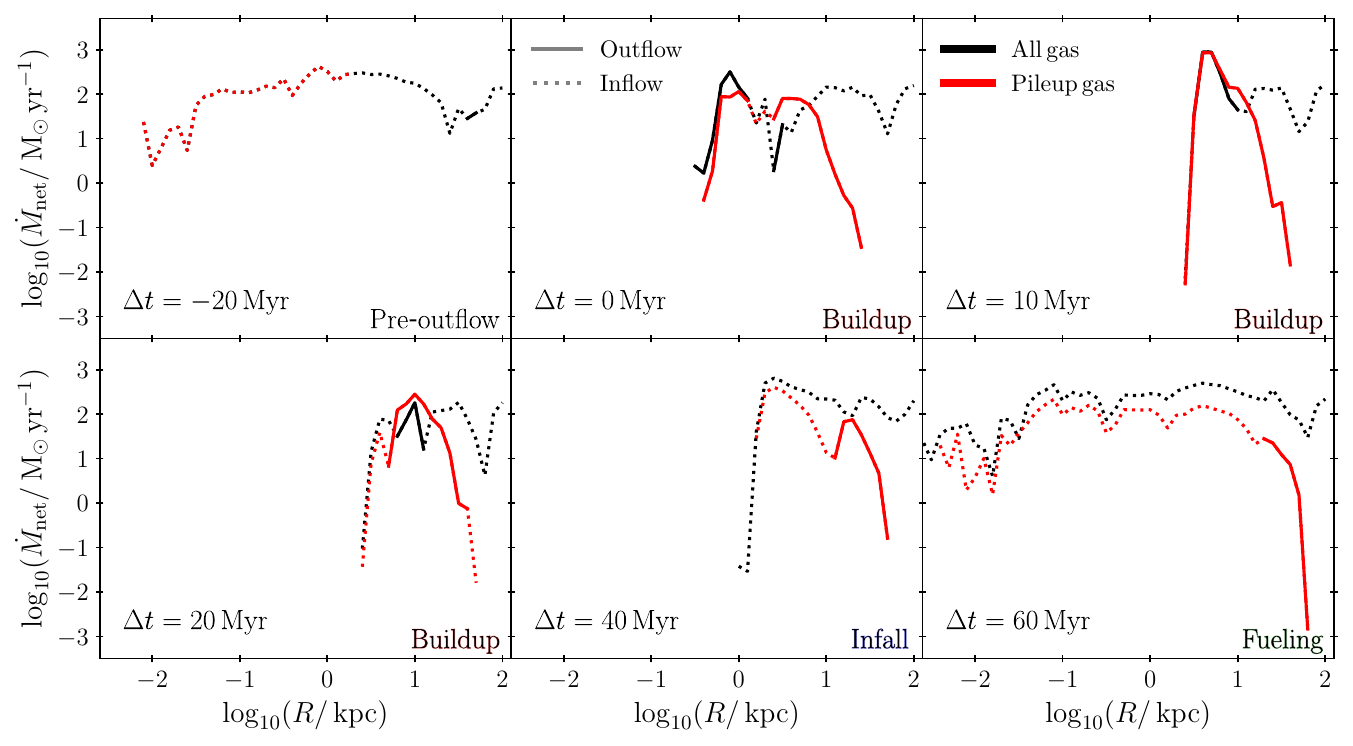}
\vspace*{-2mm}
\caption{Radial profiles similar to those in Figure~\ref{fig:decomposed_densityradprofiles}, but now displaying the net mass flow rate for all gas (black) and the contribution from pileup gas alone (shown in red for $\beta =0.5$, i.e. gas experiencing $>50$\% change in radial velocity due to the wind interaction). Net inflow is represented by dotted lines and net outflow by solid lines, with values plotted on a logarithmic scale. The net mass outflow rate increases during the first $\sim$10\,Myr up to $\sim$$1000\,\Msunyr$ as the wind propagates outward pushing the predominantly inflowing gas in the inner CGM. Re-accretion of pileup gas alone can provide $\sim$$100\, \Msunyr$ of inflowing gas down to $\lesssim$100\,pc at later times.}
\label{fig:decomposed_flowradprofiles} 
\end{figure*}

The PReS algorithm captures the propagation of the outflow in distinct stages. All gas particles initially selected to trace the propagation of the outflow are confined to the central 2\,kpc region at $\Delta t=-20$\,Myr, before the actual outflow even takes place. By $\Delta t=0$, the outflow has evacuated most gas from the central 1\,kpc region, with an expanding co-planar gas ring containing the densest pileup gas and the impact of the outflow quickly reaching larger scales. The edge-on view of the central 10\,kpc at $\Delta t=0$ clearly shows the faster expansion of gas along paths of least resistance, reaching larger scales when moving perpendicular to the galaxy plane compared to the slower expansion of the co-planar gas ring. Nonetheless, the dense gas ring continues to expand and drag surrounding gas until $\Delta t \sim 20$\,Myr, when the effective cavity size reaches $R_{\rm cavity}\sim 4$\,kpc (see Figure~\ref{fig:cavity_vs_time}). After this point, the galaxy reforms in large part by (re)accreting pileup gas previously affected by the outflow.

While the effective cavity size only reaches $\sim$4\,kpc, defined as the distance where the radial gas density profile reaches its maximum, the pileup of gas impacted by the outflow reaches much larger scales. For gas undergoing $>$50\% change in radial velocity owing to the interaction with the outflow ($\beta=0.5$), we find that pileup gas can extend out to $>$25\,kpc, well into CGM scales. The majority of gas contributing to the pileup is impacted during the first $\sim$25\,Myr of rapid expansion, but fast outflowing gas can impact the CGM on large scales even after reaching the {\it Infall} phase dominated by gas inflow onto the galaxy $\Delta t >$20\,Myr. Our analysis shows that although the majority of gas ($>$95\%) affected by the outflow lies within 25\,kpc, fast outflowing gas can extend beyond this region and continue to interact with CGM gas on larger scales, where reaccreting back into the galaxy becomes less likely. As the galaxy reforms, more than 90\% of the gas within 10\,kpc has been affected by the pileup effect, with the majority of it residing in the central 2\,kpc. This includes a substantial amount of gas that was not part of the initial outflow but was dragged before inflowing back into the galaxy.

Figure~\ref{fig:PReS_trajectories} illustrates the different types of interactions and trajectories that pileup gas can experience before contributing to the reformation of the galaxy in the {\it Fueling} phase. We focus on three gas particles as representative examples of the three types of behaviour that the PReS algorithm is designed to identify. These particles are selected to have reached the central 100\,pc region of the galaxy by $\Delta t = 60$\,Myr. On the left panel, we show the time evolution of radial velocity (top) and radial distance (bottom) for the three selected particles, which are colour-coded according to the type of interaction: Propel (P; red) indicates a particle experiencing an increase in outflowing velocity; Reversal (Re; green) indicates a particle flipping its velocity direction, from inflowing to outflowing; and Slowdown (S; purple) indicates an inflowing particle that is slowing down with a decrease in inflowing velocity. The triangle markers in the figure indicate the time when the gas particles are flagged, while the square markers denote the interaction time identified by the PReS algorithm. In the radial velocity plot, the dotted horizontal line separates inflow and outflow velocities, with positive values indicating outflowing gas and negative values indicating inflowing gas. In the radial distance plot, the grey dashed horizontal line indicates the stalling radius $R_{\rm stall}$, while the black dash-dotted line is the cavity radius $R_{\rm cavity}$ as shown in Figure~\ref{fig:cavity_vs_time}. Each particle's trajectory is represented by a line of the same colour in both plots, comparing how their radial distance and velocity change over time.

While these example particles are all selected to have reached the inner 100\,pc of the galaxy by $\Delta t=60$\,Myr, they had very different locations at $\Delta t=0$\,Myr. The Propel particle is part of the ISM of the galaxy at $R<$1\,kpc before the outflow event, subsequently ejected with peak radial velocity $V_{r}>300\,\kms$. The Reversal particle is located at $R\sim 6$\,kpc and initially inflowing towards the galaxy at $\Delta t=0$\,Myr. The Slowdown particle was also initially inflowing but located further into the CGM ($R>$10\,kpc) before interacting with the outflow, as indicated by the triangle and square markers. Interestingly, despite their rather different origins and impact from the outflow, the three particles converge to $V_{r}\sim 0\,\kms$ at a similar distance ($R\sim 5$\,kpc), illustrating the creation of a pileup of gas in the inner CGM and the subsequent coherent accretion of pileup gas onto the galaxy.

The middle and right panels of Figure~\ref{fig:PReS_trajectories} provide complementary views of the evolution of the same three gas particles, showing the 2D projected gas mass surface density distribution of the central 8\,kpc at $\Delta t=60$\,Myr with the three particles trajectories overplotted (representing their path from $\Delta t= 0\rightarrow 60$\,Myr). The grey colour bar indicates the gas mass surface density, while the particle trajectories are colour-coded to denote time evolution (middle panel) and radial velocity (right panel), The values of $\Delta t$ and $V_{r}$ are indicated by the corresponding colour bars, with time evolving from $\Delta t=0$\,Myr (blue) to $\Delta t=60$\,Myr (red) and inflowing/outflowing gas indicated in blue/red respectively. Each particle is represented by the circle marker with the same color as the left panel, denoting their initial position at $\Delta t=0$\,Myr.

The Propel particle moves outwards from the galactic centre, slowing down as it reaches beyond the stalling radius before turning back and accelerating while reaccreting onto the galaxy. The Slowdown and Reversal particles are initially inflowing until affected by the outflow. The Reversal particle is dragged/pushed back by the outflow before reaching $R_{\rm stall}$, accelerating outwards for $\sim$5\,Myr and then slowing down for $\sim$10\,Myr before reaccreting back into the galaxy. The Slowdown particle interacts with the outflow further out, decreasing its inflow velocity until becoming nearly at rest for $\sim$10\,Myr relative to the galaxy, and subsequently accelerating inward as part of the inflow reforming the galaxy. These trajectories illustrate the distinct behaviour of gas as it interacts with the outflow, highlighting the complexity of inflow-outflow dynamics. The PReS algorithm is able to identify these interactions, providing valuable insights into the overall impact of galactic outflows and the creation of a pileup of gas in the CGM that can later (re)accrete coherently in time and suddenly reform the central galaxy.

\section{Pileup Gas Fueling Galactic Nuclei} \label{sec:flagged_properties}
Figure~\ref{fig:decomposed_densityradprofiles} shows gas density as a function of radial distance, similar to Figure~\ref{fig:cavity_vs_time} but comparing the total gas distribution (black) to the contribution of pileup gas impacted by the outflow. Different colours correspond to different velocity thresholds ($\beta$ values; see \S\ref{subsec:track_methods}) to identify pileup gas, from $>$10\% change in radial velocity ($\Delta V_{r} > 0.1\,|V_{r}|$; purple) to more than a factor of four change in radial velocity owing to hydrodynamic interactions with the expanding outflow ($\Delta V_{r} > 4\, |V_{r}|$; red). The top left panel represents the initial pre-outflow conditions at $\Delta t=-20$\,Myr, where the particles selected as part of the future outflow are confined to the central 2\,kpc. The subsequent panels trace how the outflow evolves over time (corresponding to the same time snapshots as Figure~\ref{fig:flagged_maps}) and quantify the build up of pileup gas as the outflow expands and drags further gas into the CGM. The radial profiles at $\Delta t=0-20$\,Myr demonstrate the algorithm’s effectiveness at identifying outflow-affected gas. As the outflow propagates, the central few kpc are quickly cleared out, dragging infalling gas along the way, with gas piling up on scales $R\sim 4$--10\,kpc. Notably, particles identified as pileup gas closely track the ``bump'' of gas accumulated at $R\sim 4$--10\,kpc, capturing $\sim$70\% of the mass contained in this region for gas with radial velocity impacted by $>$50\% of its current velocity at the time of interacting with the outflow ($\beta \equiv \Delta V_{r}/V_{r} = 0.5$; blue). This ratio increases to $>$80\% in the subsequent panel ($\Delta t=40$\,Myr), as gas from the central region propagates further outward compressing the inflowing gas. As expected, the contribution of pileup gas to the total density distribution depends on the threshold change in radial velocity used to identify pileup gas. Gas impacted by $>$10\% of their radial velocity ($\beta=0.1$; purple) can account for $>$75\% of the mass in the central 10\,kpc at $\Delta t=40$\,Myr, while requiring a factor of four change in radial velocity ($\beta =4$; red) reduces this fraction of pileup gas to 25\%, since a lower fraction of gas is impacted so strongly by the outflow. Once the galaxy begins to reform after 40\,Myr, we see more significant differences between the total gas distribution and the pileup gas profiles in the inner 5\,kpc. In the central 1\,kpc, the average ratio of pileup to total gas density can range from 70\% to 30\%, depending on radial velocity threshold. Pileup gas is thus still a significant component after the galaxy reforms.

\begin{figure}
\includegraphics[width = \columnwidth]{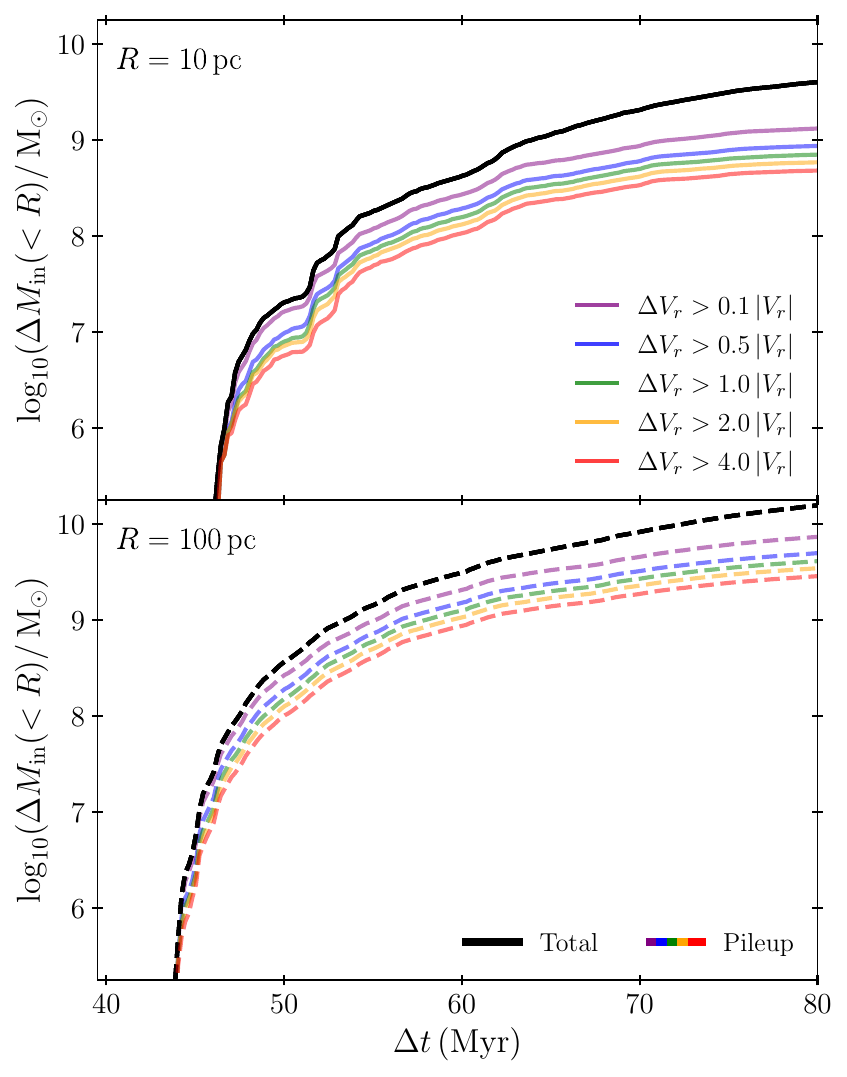}
\vspace*{-5mm}
\caption{Cumulative mass of inflowing gas across spherical shells at radius $R=10$\,pc (top: solid lines) and at $R=100$\,pc (bottom; dashed lines) as a function of time. Black lines show the total cumulative mass for all inflowing gas while colour lines are for pileup gas alone, corresponding to the different threshold changes in radial velocity owing to the wind interaction. The galaxy reforms after $\sim$45\,Myr after the global galactic wind event, with the cumulative inflow from pileup gas alone rapidly reaching $\sim$$10^{8}\,\Msun$ at $R=10\,{\rm pc}$ and $\sim$$10^{9}\,\Msun$ at $R=100\,{\rm pc}$ over the next 15\,Myr.
}
\label{fig:cumulative_mass_flow} 
\end{figure}

Figure~\ref{fig:decomposed_flowradprofiles} is similar to Figure~\ref{fig:decomposed_densityradprofiles} but quantifies the contribution of pileup gas (red; for $\beta=0.5$ in this case) to the total inflow/outflow rate (black) as a function of radial distance. For each snapshot, we compute the net mass flow rate across spherical shells of width $\Delta\log_{10}(R) = 0.1\,{\rm dex}$, depicted by dotted or solid lines for net inflow and net outflow rate, respectively. At $\Delta t=-20$\,Myr, gas is predominantly inflowing across the full range of scales (10\,pc$<R<$100\,kpc), fueling the strong starburst that eventually powers the outflow event. At $\Delta t=0$\,Myr, there is net outflow in the inner kpc, as the cavity expands, while the inflow rate still dominates on scales $R \gtrsim$3\,kpc. As expected, pileup gas tracked by the PReS algorithm is predominantly outflowing even into scales $R\sim$10\,kpc that are still dominated by net inflow, further confirming the robustness of the particle tracking algorithm. By $\Delta t=10$\,Myr, the outflow rate reaches $\sim$1000\,$\Msunyr$ at $R\sim 4$\,kpc, dominated by pileup gas. Once we reach the end of the {\it Buildup} phase ($\Delta t \sim$20\,Myr), the inner gas component begins to inflow back into the galaxy while there is still net outflow rate $>$100\,$\Msunyr$ on 10\,kpc scales. During the {\it Infall} and {\it Fueling} phases ($\Delta t >$20\,Myr), the inflow component dominates across scales, reaching a few 100\,$\Msunyr$ of inflow rate down to $<$100\,pc. We note that even after the galaxy begins accreting gas from the CGM ($\Delta t=40$--60\,Myr) pileup gas can still be seen outflowing on scales of tens of kpc (see also Figure~\ref{fig:flagged_maps}). The extent of pileup gas continues to expand owing to the propagation of fast winds along paths of least resistance. Although it represents a small fraction of gas and inflow rate on CGM scales after the reformation of the galaxy, pileup gas can contribute a significant amount of gas fueling the nuclear region of the galaxy.

Next, we investigate the amount of gas that has reached the central region of the galaxy after reformation. In Figure~\ref{fig:cumulative_mass_flow}, we quantify the contribution of pileup gas to determine the impact of the galaxy-scale outflow event fueling the nuclear region. The two panels show the cumulative mass of inflowing gas across spherical shells at $R=10$\,pc (top; solid lines) and at $R=100$\,pc (bottom; dashed lines) as a function of time, for all gas (black) and for pileup gas only, considering different $\beta$ velocity thresholds (coloured lines). The chosen scales of $R=10$\,pc and $R=100$\,pc are the smallest resolved scales in our simulation where gas is dense enough to form stars or be directly accreted by the central BH. We vary $\beta$ from 0.1 (purple) to 4 (red), where smaller $\beta$ values represent a less conservative threshold (flagging pileup particles experiencing smaller changes in radial velocity), and larger $\beta$ values require progressively greater radial velocity kicks before a particle is considered part of the pileup. We use spherical radial bins to compute the inflow rate of gas at either 10 or 100\,pc, for all gas and pileup gas. The cumulative mass over time $\Delta M_{\rm in}$ is then calculated by integrating the inflow rate over time to approximate the accumulation of gas crossing each boundary.

We see that although the outflow stalls after $\Delta t\sim 20$\,Myr (see Figure~\ref{fig:cavity_vs_time}), the galaxy remains devoid of gas for another $\sim$20\,Myr, as it takes about the same amount of time for the galaxy to reform after the outflow stalls. It then takes an additional $\sim$5\,Myr for re-accreted gas to reach the central 100\,pc region, with pileup gas dominating this early refueling. This behaviour holds for all values of $\beta$, although the amount of mass contributed by pileup gas decreases for higher $\beta$ values, as expected. While a similar trend appears at $R$\,$=$\,10\,pc, the cumulative amount of gas flowing across the 10\,pc boundary after $\Delta t=80$\,Myr is about one order of magnitude lower than that at $R=100$\,pc. This is expected, as there is gas consumption by star formation and only a fraction of gas on 100\,pc scales can lose enough angular momentum to reach down to $<$10\,pc \citep{Angles-Alcazar2021}. Overall, pileup gas contributes $>$$10^{8}\,\Msun$ of cumulative inflowing gas to $<$10\,pc in the first $\sim$10\,Myr after reformation of the galaxy.

\begin{figure*}
\includegraphics[width = \textwidth]{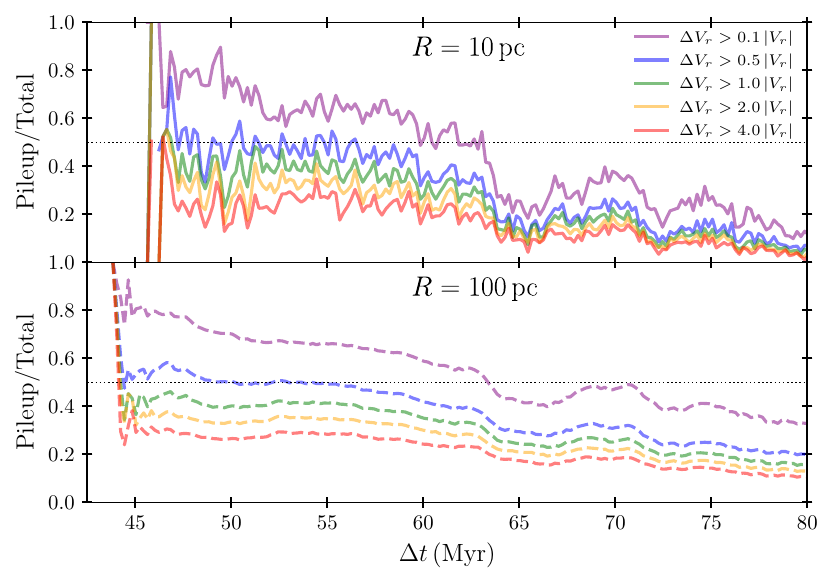}
\vspace*{-5mm}
\caption{Fraction of mass from pileup gas relative to the total gas mass within a sphere of either 10\,pc (top; solid lines) or 100\,pc (bottom; dashed lines) around the BH. Each colour represents a different radial velocity threshold, from >10\% change in radial velocity ($\Delta V_{r} > 0.1\,|V_{r}|$; purple) to more than a factor of four change in radial velocity ($\Delta V_{r} > 4\,|V_{r}|$; red). The dotted horizontal line denotes 50\%. Pileup gas can contribute as much as 80\% of the nuclear gas reservoir soon after the reformation of the galaxy, with gas strongly impacted by the wind interaction ($\Delta V_{r} > 4\,|V_{r}|$) contributing $\sim$30\%.  Pileup gas is most important in the first $\sim$15\,Myr, with its contribution declining at later times.}
\label{fig:massfraction} 
\end{figure*}

Figure~\ref{fig:massfraction} presents the \emph{pileup fraction}, defined as the mass of outflow-affected (pileup) gas within a spherical volume of radius \(R\) divided by the total gas mass within that same radius, shown as a function of time. The top and bottom panels display results for \(R=10\,\mathrm{pc}\) and \(R=100\,\mathrm{pc}\)), respectively, with colour-coded lines indicating various values of \(\beta\). The dotted horizontal line marks the 50\% threshold, highlighting the point at which half of the gas mass in that region comes from pileup gas. Unlike the cumulative inflowing mass shown in Figure~\ref{fig:cumulative_mass_flow} which tracks the mass of gas crossing a given radial shell, the pileup fraction shown here measures the instantaneous fraction of mass already present within the specified volume that was previously affected by the outflow. 

Once gas has reached the nuclear region of the galaxy at $\Delta t \sim 45$\,Myr, we find that pileup gas for $\beta$\,$=$\,0.5 can contribute $\sim$50\% of the existing gas mass in the central 100\,pc over the subsequent $\sim$15\,Myr. This implies that a significant fraction of the nuclear gas supply right after reformation of the galaxy can be traced back to gas that experienced $>$50\% change in radial velocity owing to the wind interaction. The fraction of pileup gas in the inner 100\,pc decreases over time, but remains $\sim$20\% after $\Delta t=80$\,Myr for $\beta=0.5$. As seen in Figure~\ref{fig:cumulative_mass_flow}, the amount of pileup gas depends on the strength of the inflow-outflow interaction, with pileup gas contributing 80\%\,$\rightarrow$\,30\% for $\beta=0.1$ and 30\%\,$\rightarrow$\,10\% for $\beta$\,$=$\,4 during the time interval $\Delta t=45\,{\rm Myr}$\,$\rightarrow$\,$80\,{\rm Myr}$. We find similar results for the fraction of pileup gas in the inner 10\,pc region, indicating stronger effects during the first 20\,Myr and a declining contribution at later times. Overall, these results suggest that the sudden (re)accretion of pileup gas can represent a crucial nuclear fueling mode for galaxies at their peak of activity.

\begin{figure}
\includegraphics[width = \columnwidth]{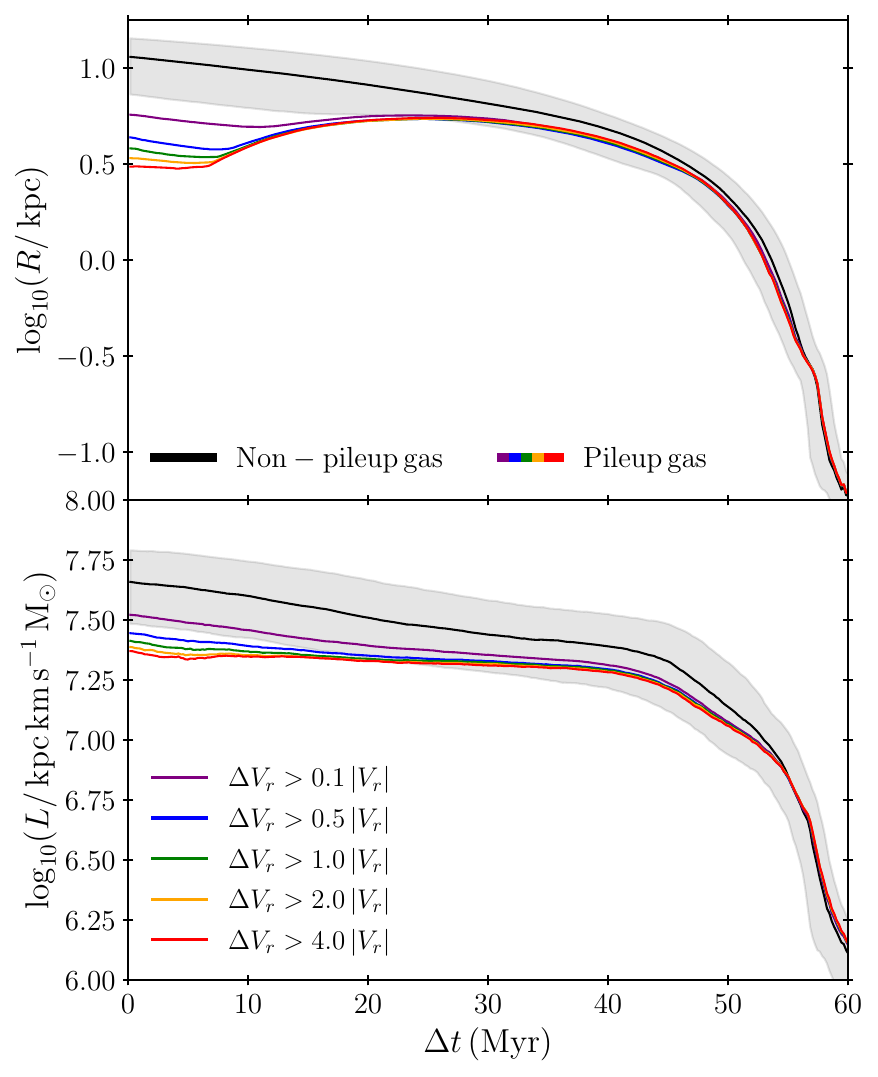}
\vspace*{-5mm}
\caption{Time evolution of radial distance (top) and angular momentum (bottom) for gas selected to have reached the inner 100\,pc of the galaxy at $\Delta t=60\,{\rm Myr}$. Gas not impacted by the galactic wind pileup effect (for $\beta=4$) is shown in black, with the solid line and shaded regions indicating the median and 25$^{\rm th}$-75$^{\rm th}$ percentile range. Coloured lines represent median trends for pileup gas identified for different thresholds in radial velocity change. Pileup gas originates closer to the galaxy and has correspondingly lower angular momentum at $\Delta t=0$\,Myr than non-pileup gas, but the expanding outflow has no systematic effect on the angular momentum evolution of pileup gas.
}
\label{fig:within_100pc} 
\end{figure}

\section{Discussion} \label{sec:discussion}

Central starbursts and massive BHs are thought to be fueled by gas funneled to the nuclear region of the galaxy through two main mechanisms: (i) major mergers of gas-rich galaxies, where tidal forces efficiently drive large amounts of gas into the central regions \citep{Canalizo2001,Veilleux2002,Springel2005,Hopkins2006,Engel2010,Treister2012,Cotini2013,Blecha2018,Blumenthal2018,Diaz-Santos2018,Ellison2019}; and (ii) minor mergers and secular processes in gravitationally unstable disks, including stellar bars and spiral arms, which can steadily channel gas inward without the need for major external perturbations \citep{Kormendy2004,Garcia-Burillo2005,Hopkins&Quataert2010,Kaviraj2014,Angles-Alcazar2021,Combes2023}. The relative importance of these two fueling mechanisms for BH growth remains a subject of ongoing debate. Observationally, some studies report only weak correlations between mergers and AGN activity at high redshifts~\citep{Pierce2007,Cisternas2011,Kocevski2012,Villforth2014,Villforth2019,Villforth2023}. A similar tension exists in theoretical models. While idealized galaxy merger simulations invariably show that major mergers trigger AGN \citep[e.g.,][]{Hopkins&Quataert2010,Pontzen2017}, this connection is more complex and not as direct in cosmological simulations \citep{Ricarte2019,McAlpine2020,Sharma2024}, where minor mergers are more common \citep{Byrne-Mamahit2025}.
Resolving this tension requires a deeper understanding of the mechanisms that can transport gas across many orders of magnitude in spatial scales to ultimately fuel the central BH.

In this work, we have proposed a new mechanism to deliver a substantial amount of gas down to the nuclear region of galaxies over a short timescale, complementing major galaxy mergers and secular processes as primary fueling channels. The pileup gas delivery mechanism presented here relies on the global impact of galactic winds driven by stellar feedback. Large-scale outflows evacuate the ISM content of galaxies and deposit a large amount of gas in the inner CGM, efficiently dragging gas while expanding and slowing down accretion from the CGM. For a massive ($M_{\star}\sim10^{10.5}\,\Msun$), star-forming galaxy at its peak of activity ($z \gtrsim 2$), this can create a $\sim$$10^{10}\,\Msun$ pileup of gas in the inner CGM subsequently accreting onto the galaxy as a coherent inflow event. Using FIRE cosmological simulations coupled with a novel particle tracking algorithm, we have shown that this galactic wind pileup mechanism can achieve a tenfold increase in inflow rate over pre-outflow conditions, providing a dominant contribution to the gas supply in the central $\sim$10--100\,pc. While galaxy interactions and gravitational instabilities can trigger the initial star formation-driven outflow, the proposed galactic wind pileup effect may represent a significant fueling mode for compact starbursts and luminous quasars in massive galaxies at the peak of activity.

Previous studies have investigated the role of stellar mass return in sustaining late time star formation and AGN \citep[e.g.,][]{Leitner2011,Segers2016,Salvador-Rusinol2021,Benedetti2023,Riffel2023}. For instance, \citet{Ciotti2007} proposed that gas returned to the ISM by evolved stars can trigger both central starbursts and AGN activity in simulations of isolated elliptical galaxies, acting as the primary fuel source in systems dominated by aging stellar populations. 
In more recent work, \citet{Choi2024} used cosmological simulations including BH growth and feedback \citep{Choi2017,Choi2018} to investigate the origin of gas feeding BHs with particle tracking. They found that feedback-ejected gas—i.e., wind-recycled material—was the dominant source, comprising $\sim$40\% of the accreted mass while also contributing, in a lesser extent, to star formation in their simulations. 
Our simulations also implement mass return from stellar evolution, including Type II and Type Ia SNe, mass loss from OB stars, and AGB winds, but we do not explicitly quantify here their contribution to nuclear fueling. Instead, we evaluate the impact of a global galactic outflow on regulating the galaxy gas supply from the CGM and causally driving significantly higher gas inflow rates once the galaxy reforms at later times. We find that gas flagged as pileup during the $\sim$40\,Myr outflow period contributes over 50\% of the material to fuel the central region. We also show that within the first $\sim$10\,Myr of the {\it Fueling} phase, over 60\% of the material reaching the central 100\,pc—and ultimately the innermost 10\,pc—is pileup gas. This demonstrates that feedback-processed gas can be dominant and immediate fuel source for nuclear activity.

Regardless of the origin of gas, angular momentum is known to represent a major barrier to BH fueling: gas must lose $>$99\% of its angular momentum to feed the accretion disc \citep{Jogee2006,AlexanderHickox2012,Alexander2025}. Idealized multi-scale galaxy merger simulations \citep{Hopkins&Quataert2010,Hopkins&Quataert2011} and cosmological hyper-refinement simulations \citep{Angles-Alcazar2021,Hopkins2024_zoom1} have shown that stellar gravitational torques dominate angular momentum transport from galaxy scales down to sub-pc scales and can feed luminous quasar outflows in gas-rich discs under the presence of strong non-axysymmetric perturbations to the stellar potential, either from galaxy mergers or secular gravitational instabilities. Our proposed galactic wind pileup fueling mode complements this picture while still relying on angular momentum transport by traditional mechanisms. The pileup mechanism simply acts as a transient gas reservoir, without necessarily impacting angular momentum, accumulating gas in the inner CGM during the outflow {\it Buildup} phase and then suddenly releasing a coherent gas inflow providing a significantly higher accretion rate than average conditions.

To test whether pileup gas experiences any systematic change in angular momentum we analysed the prior evolution of gas particles that had reached the inner 100\,pc shortly after the onset of the {\it Fueling} phase ($\Delta t=60\,{\rm Myr}$). Figure~\ref{fig:within_100pc} presents the median evolution of these particles' radial distance (top) and angular momentum (bottom) as they are tracked backward in time. We identify which of these particles had previously been identified as part of the pileup of gas in the inner CGM, using varying thresholds $\beta$ for the required change in radial velocity ($\Delta V_{r} > \beta\,|V_{r}|$). Coloured lines represent the median distance and angular momentum evolution of pileup gas identified for different velocity thresholds. For comparison, we define non-pileup gas as those particles that were never identified as pileup under the most conservative threshold ($\beta=4$); this group is shown as the black line, with the shaded region indicating the 25$^{\rm th}$–75$^{\rm th}$ percentile range.

As expected, we find that pileup gas originates closer to the galaxy's centre and has correspondingly lower initial angular momentum than the non-pileup gas that also inflows down to $<$100\,pc at $\Delta t= 60$\,Myr. Furthermore, we see systematic differences in radial trajectories for pileup gas with different $\beta$ thresholds, with the gas most impacted by the outflow ($\beta=4$; red) originating closer to the galaxy. Interestingly, all median pileup gas trajectories show clear outward motion due to the impact of the expanding outflow, reaching $\sim$6\,kpc at $\Delta t$\,$\sim$\,20--30\,Myr across all velocity thresholds before inflowing down to the nuclear region. This consistent turnaround point signifies the large pileup of gas created in the inner CGM and suggests a characteristic delay time by the outflow before gas can globally accrete again. Nonetheless, we find no clear impact of the outflow on the overall angular momentum evolution of pileup gas, generally following the progressive decrease in angular momentum seen for non-pileup gas that also accretes down to the galaxy nucleus. The pileup mechanism coordinates the (re)accretion of substantial amounts of gas over a short timescale. It modulates the timing of delivery rather than driving angular momentum loss.

Previous simulations have shown that gas ejected in galactic winds can either increase or decrease its specific angular momentum before recycling back to the galaxy, depending on the angular momentum of the CGM \citep{Christensen2016,Grand2019,Hafen2022}. Figure~\ref{fig:within_100pc} shows no indication of increased angular momentum for pileup gas, suggesting that for the case of star-forming galaxies fed primarily by cold accretion flows at early times, the gas pileup may retain a substantial amount of low specific angular momentum gas in the inner CGM, favoring efficient (re)accretion down to the centre of the galaxy. Furthermore, our findings reveal that gas expelled by feedback does not simply escape or return passively, instead it can dynamically reshape the inflow environment. Outflow-induced cavities can temporarily prevent inflows, leading to the buildup of pileup gas that later fuels a sudden burst of activity. This suggests a gas replenishment mechanism differing from classical ``fountain flows'' and wind recycling loops that consider only gas ejected from the galaxy and reaccreted back \citep{Oppenheimer2010,Christensen2016,Grand2019,Tollet2019}. In contrast, inflowing gas from the proposed pileup effect, greatly increases the amount of gas available for the subsequent inflow event. Once most of the gas that we consider to have been recycled has been exhausted, the major composition of this pileup gas becomes that which was accreted from the CGM, with a large amount of pileup gas, visible in the final $\sim$40\,Myr, now able to reach the galactic centre after being temporarily prevented by the outflow. By identifying gas affected by the outflow across the full range of scales, not just directly ejected from the galaxy but also dragged throughout the CGM, our PReS particle tracking algorithm offers a more detailed account of the global impact of outflows. This offers a powerful new approach to investigate the interplay of inflows-outflows, the distribution of mass and metals, and their impact in galaxy evolution.

In this work, we have focused on the ``last'' major galactic wind event immediately before the simulated galaxy transitions from bursty to smooth star formation and from prominent global galactic winds to inefficient stellar feedback. This transition occurs near the stellar mass scale $M_{\star}\sim 10^{10.5}\,\Msun$, where BHs begin to accrete more efficiently.     
At this stage, AGN feedback becomes critical for quenching star formation and regulating central stellar densities in massive galaxies. Once a galaxy has built a sufficiently deep stellar potential, gas can be retained and funneled onto the BH \citep{Angles-Alcazar2021,Byrne2023,Hopkins2023}. This sets the stage for rapid BH growth, alignment with the observed $M_{\rm BH}-M_{\star}$ relation at lower redshifts, and the onset of strong AGN feedback \citep{Wellons2023,Byrne2024}. 
We have shown that the pileup mechanism can deliver $>$10$^{9}\,\Msun$ of gas down to the central 10\,pc (Figure~\ref{fig:cumulative_mass_flow}), suggesting that even if only a small fraction accretes down to the central BH, pileup gas can significantly contribute to BHs converging onto the BH--galaxy relation.
As demonstrated in previous analyses of the simulated galaxy presented in this work, at $z \sim 2$ when star formation is vigorous, stellar feedback alone becomes insufficient to regulate galaxy growth. This leads to overcompact and overdense stellar components \citep{Wellons2020,Parsotan2021}, underscoring the necessity of an additional feedback mechanism, such as AGN feedback, to properly regulate galaxy evolution at this critical stage. More recent re-simulations of this galaxy including a novel implementation of hyper-refined AGN winds have shown that, under these conditions, strong AGN feedback can indeed suppress star formation provided that the BH is efficiently fueled \citep[][]{Cochrane2023a,Mercedes-Feliz2023,Mercedes-Feliz2024}. Lagrangian hyper-refinement simulations of this galaxy have confirmed explicitly that stellar gravitational torques can drive large gas inflow rates down to sub-pc scales, powering a luminous quasar phase at this critical epoch \citep{Angles-Alcazar2021}, but the origin of the galaxy gas reservoir was left unexplained. The pileup mechanism directly addresses this gap, providing crucial gas replenishment at a particularly relevant time in a galaxy’s life cycle. The last major galactic wind event drives the pileup, which then refuels the galaxy precisely when the stellar potential becomes strong enough to prevent further gas evacuation. This creates the ideal conditions for quasar fueling and sets the stage for AGN feedback to take over as the dominant regulator of central star formation.

In this context, it is natural to ask what role AGN feedback would play at this critical juncture. While stellar feedback can drive the large-scale outflow that creates the gas pileup, it is ultimately insufficient to prevent the gas fallback and sustained starburst phase observed in this simulation. As shown in previous work \citep[][]{Mercedes-Feliz2023, Mercedes-Feliz2024}, a stronger, faster outflow driven by AGN winds could work in tandem with stellar feedback, further evacuating the CGM and potentially preventing the reaccretion of gas in the spatial and time scales that we see here. 
While our simulations did not include any prior effect of AGN feedback, BH accretion is expected to be less efficient at earlier times \citep{Dubois2015,Angles-Alcazar2017c,Habouzit2017,Angles-Alcazar2021,Byrne2023} and thus we do not expect the lack of AGN feedback to significantly affect our conclusions regarding the pileup effect. 
Stellar feedback is the primary regulator earlier in the galaxy's life, with the pileup mechanism providing a crucial fueling event that sets the stage for rapid BH growth and the subsequent onset of powerful AGN feedback, ultimately serving as the dominant quenching mechanism at these scales.

Our results align with a broader body of work showing that powerful, continuously active AGN feedback can be more effective than stellar feedback at suppressing gas reaccretion, particularly in high-mass systems \citep[e.g.,][]{Choi2017, Costa2018b, Pillepich2018}.
Nonetheless, AGN wind events could drive their own pileup effect, potentially refueling the BH at later times \citep{Dehnen2013}. 
However, the fate of AGN-ejected material in simulations is highly model-dependent: different implementations of AGN feedback (thermal vs kinetic vs jet modes), halo masses, and duty cycles can produce a spectrum of outcomes. High-resolution zoom-ins such as the FIRE-3 simulations \citep{Byrne2024}, NewHorizon \citep{Dubois2021}, and RomulusC \citep{Tremmel2019}, as well as large-volume models such as SIMBA \citep{Dave2019} and IllustrisTNG50 \citep{Nelson2019} explore these regimes but do not uniformly demonstrate if gas expelled specifically by AGN winds later reaccretes. Where recycling analyses have been performed, the conclusions depend sensitively on the feedback prescription and the halo environment. Thus, rather than predicting a single outcome, previous work indicates a diversity of possibilities. Therefore mechanisms like the pileup effect proposed here may remain relevant in regimes where AGN are intermittent, weak, or otherwise inefficient at permanently evacuating the CGM. Therefore, the pileup mechanism may still play a crucial role in scenarios where AGN feedback is not continuously dominant or efficient, by transiently accumulating gas in the CGM before it is funneled back to the galaxy centre. In this way, our findings complement prior work: stellar feedback can excavate the initial cavity, but without sustained AGN winds the gas eventually reaccretes; with strong AGN feedback, however, the balance between ejection and recycling could shift, potentially delaying or even preventing fallback. 

Having established this theoretical framework, we now turn to the observational consequences. The pileup mechanism predicts a short-lived ($\sim$10--20\,Myr) ``cavity and thin shell'' morphology in the inner CGM as the stellar-feedback–driven outflow stalls and accumulates. 
The central cavity (hot, low-density gas) and the surrounding thin, dense shell (cooler, higher-density gas) should imprint distinct multi-wavelength signatures: (i) hot cavity: depressed X-ray surface brightness coincident with enhanced gas temperature (e.g., AGN-inflated cavities; \citealt{Hlavacek-Larrondo2015}); (ii) thin shell: bright, shock-enhanced nebular lines (e.g., H$\alpha$, H$\beta$, [O III]$\lambda$5007, [O II]$\lambda$3727, [S II]$\lambda\lambda$6717,6731, [O I]$\lambda$6300) with ring-like morphology, elevated densities diagnosed via [S II] and [O II] doublet ratios \citep{Sanders2016}, and split or broadened profiles tracing an expanding/contracting shell with velocity separations of a few $10^{2}\,\kms$ (expanding superbubbles; \citealt{Egorov2023}). At $z\sim2$, these lines fall in the near-IR and are accessible with JWST/NIRSpec IFU, which provides $\sim$0.1'' resolution \citep{Boker2022}, and would be readily resolved, especially in lensed systems or with ground-based adaptive optics IFUs. Cold gas re-condensed in the shell should appear as a molecular or dust ring (CO, [C I], dust continuum) at radii of $\sim$1--3\,kpc, observable with ALMA at 0.02--0.1'' resolution (e.g., HL Tau long-baseline imaging; \citealt{ALMAPartnership2015}).
JWST has already demonstrated its capability to detect such structures: in nearby galaxies, MIRI and NIRCam resolve arc-like CGM dust features down to $\sim$12\,pc scales (NGC 891; \citealt{Chastenet2024}); PHANGS–JWST observations of NGC 1365 reveal a molecular gas ring at $\sim$475\,pc radius fed by converging gas flows (\citealt{Schinnerer2023}); and JADES/NIRSpec has spatially resolved ionized-gas kinematics at 
$z > 5.5$ on $\sim$1\,kpc scales, detecting broad and structured emission lines consistent with outflowing and inflowing components \citep{deGraaff2024}. At higher redshift, JWST has also revealed extended nebular emission and bubble-like morphologies associated with AGN feedback (e.g. \citealt{Saxena2024}). We can kinematically test our model by looking for a specific temporal sequence: (1) an outward shell with line splitting and declining central ISM emission (evacuation), followed by (2) re-accretion signatures—redshifted inflow components along the major axis, increasing central gas surface density, a compact starburst (high $\Sigma_{\rm SFR}$), and potentially rising BH accretion (broad-line widths, [O III] wings, high $L_{\rm bol}/ L_{\rm Edd}$). Because the cavity/shell phase is brief, the duty cycle should be low. Detecting it therefore requires wide surveys or lens-selected targets, and stacking can enhance detectability (e.g. average X-ray depressions).

The pileup-driven fueling scenario offers a natural explanation that is consistent with JWST indications of strong early stellar feedback, including the steep faint-end of the UV luminosity function \citep{Harikane2023,Boyett2024}, the prevalence of compact or ``mini-quenched'' systems \citep{Carnall2024,Looser2024,Looser2025}, and evidence for rapid black hole growth at high redshift \citep{Kocevski2023,Ubler2023,Maiolino2024}. The pileup-driven fallback provides a plausible mechanism to drive a transient phase of rapid BH fueling by synchronizing central gas density spikes, compact starbursts, and elevated BH accretion rates. We therefore predict that galaxies caught shortly after a strong, stellar-feedback–driven outflow will show (a) compact sizes, (b) elevated central gas fractions and $\Sigma_{\rm SFR}$, (c) evidence for recent outward shell kinematics, and (d) AGN indicators with high accretion ratios, before transitioning to AGN-dominated feedback at slightly later times.

Although we have focused on the ``last'' major galactic wind, particularly effective at fueling quasars during their rapid growth phase, earlier winds may have produced qualitatively similar effects, albeit with more complex cumulative impacts. In this picture, successive wind episodes could recycle gas back to the nucleus on different timescales, extending the fueling period for BH growth and star formation. Such recurrent recycling may help explain the observed intermittency of AGN activity and the persistence of central starbursts at early times. A related mechanism operating at late times has been explored in idealized models of galaxy groups and clusters, where AGN‐driven jets trigger thermal instabilities in hot halo gas, allowing cold clumps or filaments to condense and ``rain down'' onto the central BH, fueling subsequent active episodes \citep{Li2015,Voit2015}. For example, \citet{Talbot2022} show in jet simulations that beyond merely disrupting or clearing gas, jets can induce condensation of hot gas in the CGM and produce backflows and cooling‐induced inflows that feed the BH. Together, these scenarios suggest that stellar feedback-driven winds predominantly recycle gas on galactic scales, while jet-induced precipitation governs gas cooling and fueling on larger halo or cluster scales, with both channels contributing to recurrent AGN activity.
Observational constraints suggest that quasar activity consists of numerous short episodes \citep{Schawinski2015,Eilers2017,Nyland2020,Shen2021,Eilers2025}, requiring physical mechanisms that rapidly switch BH accretion on and off. In addition to intrinsic variability in fueling supply \citep{Levine2008,Hopkins&Quataert2010,Angles-Alcazar2021}, AGN feedback has been proposed as a key mechanism driving variability \citep{Novak2011,Padovani2017}.
Our results suggest that a large fraction of pileup gas can re-enter the nuclear region within a few tens of Myr after being expelled, providing a recurrent ``fuel return'' mechanism that could explain both the high variability and short duty cycles of luminous AGN, especially when coupled with large-scale galactic outflows.

\section{Conclusions}\label{sec:conclusions}
We present a detailed analysis of the impact of the last major galactic-scale outflow driven by stellar feedback in a simulated massive galaxy near the peak of star formation activity ($M_{\rm halo} \sim 10^{12.5}\,{\rm M}_{\odot}$ at $z\sim2$) on the subsequent (re)fueling of the nuclear region and possible triggering of AGN activity. We use a novel particle tracking algorithm to follow and quantify the role that the global outflow plays on building up a gas reservoir in the inner CGM that later (re)accretes coherently onto the galaxy, driving a large amount of gas and fueling a compact starburst and/or luminous quasar phase. Our simulations include local stellar feedback and resolved multi-phase ISM physics from the FIRE-2 project \citep{Hopkins2018}. Our main results can be summarised as follows:

\begin{enumerate}[wide, labelwidth=!,itemindent=!]
    \item The stellar feedback-driven outflow evacuates gas from the central $\sim$4\,kpc and prevents further accretion onto the galaxy over a period of $\sim$20\,Myr, creating a ``pileup'' of gas of $\gtrsim$$10^{10}\,\Msun$ at the interface where the expanding outflowing material interacts with and sweeps up ambient CGM gas.

    \item After the outflow stalls, the pileup of gas in the inner CGM re-accretes coherently onto the galaxy, driving extreme gas inflow rates reaching $\sim$100--1000\,$\Msunyr$ at $<$100\,pc and 10--100\,$\Msunyr$ at $<$10 pc and central gas densities $\gtrsim$$10^{11}\,\Msun\,{\rm kpc}^{-3}$. Despite the intense SFR (see Figure 1 in \citealt{Angles-Alcazar2021}), stellar feedback can no longer evacuate gas from the nuclear region owing to the deepening stellar gravitational potential, creating the ideal conditions for fueling the central massive BH.

    \item The increased mass inflow rate and central gas density remain elevated for $\gtrsim$40\,Myr and significantly higher than the pre-outflow conditions, suggesting that the outflow event is not merely regulating galaxy growth but a catalyst for the subsequent reformation of the galaxy and sudden increase in the nuclear fueling rate.
    
    \item We have developed and applied the PReS particle tracking algorithm to identify gas that is either propelled in galactic winds, reversed from inflowing to outflowing by gas dragging on CGM scales, or slowed down while accreting from the CGM. By explicitly tracking changes in radial velocity owing to hydrodynamic interactions with outflowing material, this novel methodology enables the identification of ``pileup'' particles significantly impacted by the outflow event and their contribution to the subsequent fueling of the nuclear region.

    \item Following the origin and fate of gas that reaches the central 10--100\,pc of the galaxy over time, we find that pileup gas can contribute $>$50\% of the nuclear gas reservoir soon after the reformation of the galaxy, linking the immediate aftermath of the outflow to central fueling. Gas strongly impacted by the wind interaction (with more than a factor of four change in radial velocity) contributes $\sim$30\% of nuclear gas and remains a significant fueling component during the first $\sim$15\,Myr.

    \item The pileup effect does not systematically impact the angular momentum of gas that ends up fueling the nuclear region, with stellar gravitational torques dominating angular momentum transport on galaxy scales.  Instead, the galactic wind pileup mechanism proposed here simply accumulates and then suddenly releases a substantial amount of gas that would otherwise slowly accrete onto the galaxy, driving extreme inflow rates over a short period of time.

    \item The pileup of gas driven by the last global galactic wind powered by star formation provides the ideal conditions for quasar fueling, when the deepening stellar potential prevents further gas evacuation by stellar feedback and AGN feedback is most needed to regulate the growth of massive galaxies at their peak of activity.

\end{enumerate}
These results demonstrate a causal connection between a major galactic outflow and the subsequent fueling of the nuclear gas reservoir.  While we focus on the last global star formation-driven outflow in the formation history of a massive galaxy, earlier galactic winds and/or AGN-driven outflows could have qualitatively similar effects, creating a pileup of gas in the CGM that can later accrete coherently onto the galaxy and play a role in gas dynamics and star formation cycles.    
Future work will expand the investigation into the role of this galactic wind pileup fueling mechanism in simulations including explicitly BH growth and feedback. Combining diagnostics of angular momentum transport with our pileup-tracking method will be crucial to understanding how different gas supply modes shape BH feeding and feedback cycles across cosmic time. In parallel with theoretical efforts, future work should investigate plausible observational signatures of the proposed pileup effect, exploring connections between CGM properties, starburst signatures, and statistical properties of AGN hosts.

\section*{Acknowledgements}
The simulations were run on Flatiron Institute’s research computing facilities, supported by the Simons Foundation. Additional numerical calculations were run on the Caltech compute cluster “Wheeler,” allocations FTA-Hopkins supported by the NSF and TACC, and NASA HEC SMD-16-7592, and XSEDE allocation TG-AST160048 supported by NSF grant ACI-1053575.
JMF was supported in part by a NASA CT Space Grant.
DAA acknowledges support from NSF grant AST-2009687 and CAREER award AST-2442788, NASA grant ATP23-0156, STScI JWST grants GO-01712.009-A, AR-04357.001-A, and AR-05366.005-A, an Alfred P. Sloan Research Fellowship, and Cottrell Scholar Award CS-CSA-2023-028 by the Research Corporation for Science Advancement.
RKC is grateful for support from the Leverhulme Trust via the Leverhulme Early Career Fellowship.
SW received support from NASA grant 80NSSC24K0838.
CAFG was supported by NSF through grants AST-2108230 and AST-2307327; by NASA through grants 21-ATP21-0036 and 23-ATP23-0008; and by STScI through grant JWST-AR-03252.001-A.

\section*{Data Availability}
The data supporting the plots within this article are available on reasonable request to the corresponding author. FIRE-2 simulations are publicly available \citep{Wetzel2023,Wetzel2025} at \url{http://flathub.flatironinstitute.org/fire}. Additional FIRE simulation data, including initial conditions and derived data products, are available at \url{https://fire.northwestern.edu/data/}. A public version of the GIZMO code is available at \url{http://www.tapir.caltech.edu/~phopkins/Site/GIZMO.html}.



\bibliographystyle{mnras}
\bibliography{main} 

\begin{thebibliography}{}
\makeatletter
\relax
\def\mn@urlcharsother{\let\do\@makeother \do\$\do\&\do\#\do\^\do\_\do\%\do\~}
\def\mn@doi{\begingroup\mn@urlcharsother \@ifnextchar [ {\mn@doi@}
  {\mn@doi@[]}}
\def\mn@doi@[#1]#2{\def\@tempa{#1}\ifx\@tempa\@empty \href
  {http://dx.doi.org/#2} {doi:#2}\else \href {http://dx.doi.org/#2} {#1}\fi
  \endgroup}
\def\mn@eprint#1#2{\mn@eprint@#1:#2::\@nil}
\def\mn@eprint@arXiv#1{\href {http://arxiv.org/abs/#1} {{\tt arXiv:#1}}}
\def\mn@eprint@dblp#1{\href {http://dblp.uni-trier.de/rec/bibtex/#1.xml}
  {dblp:#1}}
\def\mn@eprint@#1:#2:#3:#4\@nil{\def\@tempa {#1}\def\@tempb {#2}\def\@tempc
  {#3}\ifx \@tempc \@empty \let \@tempc \@tempb \let \@tempb \@tempa \fi \ifx
  \@tempb \@empty \def\@tempb {arXiv}\fi \@ifundefined
  {mn@eprint@\@tempb}{\@tempb:\@tempc}{\expandafter \expandafter \csname
  mn@eprint@\@tempb\endcsname \expandafter{\@tempc}}}

\bibitem[\protect\citeauthoryear{{ALMA Partnership} et~al.,}{{ALMA Partnership}
  et~al.}{2015}]{ALMAPartnership2015}
{ALMA Partnership} et~al., 2015, \mn@doi [\apjl] {10.1088/2041-8205/808/1/L3},
  \href {https://ui.adsabs.harvard.edu/abs/2015ApJ...808L...3A} {808, L3}

\bibitem[\protect\citeauthoryear{{Agertz} \& {Kravtsov}}{{Agertz} \&
  {Kravtsov}}{2016}]{Agertz2016}
{Agertz} O.,  {Kravtsov} A.~V.,  2016, \mn@doi [\apj]
  {10.3847/0004-637X/824/2/79}, \href
  {https://ui.adsabs.harvard.edu/abs/2016ApJ...824...79A} {824, 79}

\bibitem[\protect\citeauthoryear{{Agertz} et~al.,}{{Agertz}
  et~al.}{2020}]{Agertz2020}
{Agertz} O.,  et~al., 2020, \mn@doi [\mnras] {10.1093/mnras/stz3053}, \href
  {https://ui.adsabs.harvard.edu/abs/2020MNRAS.491.1656A} {491, 1656}

\bibitem[\protect\citeauthoryear{{Alexander} \& {Hickox}}{{Alexander} \&
  {Hickox}}{2012}]{AlexanderHickox2012}
{Alexander} D.~M.,  {Hickox} R.~C.,  2012, \mn@doi [\nar]
  {10.1016/j.newar.2011.11.003}, \href
  {https://ui.adsabs.harvard.edu/abs/2012NewAR..56...93A} {56, 93}

\bibitem[\protect\citeauthoryear{{Alexander} et~al.,}{{Alexander}
  et~al.}{2025}]{Alexander2025}
{Alexander} D.~M.,  et~al., 2025, \mn@doi [\nar] {10.1016/j.newar.2025.101733},
  \href {https://ui.adsabs.harvard.edu/abs/2025NewAR.10101733A} {101, 101733}

\bibitem[\protect\citeauthoryear{{Andersson}, {Agertz}, {Renaud}  \&
  {Teyssier}}{{Andersson} et~al.}{2023}]{Andersson2023}
{Andersson} E.~P.,  {Agertz} O.,  {Renaud} F.,   {Teyssier} R.,  2023, \mn@doi
  [\mnras] {10.1093/mnras/stad692}, \href
  {https://ui.adsabs.harvard.edu/abs/2023MNRAS.521.2196A} {521, 2196}

\bibitem[\protect\citeauthoryear{{Angl{\'e}s-Alc{\'a}zar}, {{\"O}zel}  \&
  {Dav{\'e}}}{{Angl{\'e}s-Alc{\'a}zar} et~al.}{2013}]{Angles-Alcazar2013}
{Angl{\'e}s-Alc{\'a}zar} D.,  {{\"O}zel} F.,   {Dav{\'e}} R.,  2013, \mn@doi
  [\apj] {10.1088/0004-637X/770/1/5}, \href
  {https://ui.adsabs.harvard.edu/abs/2013ApJ...770....5A} {770, 5}

\bibitem[\protect\citeauthoryear{{Angl{\'e}s-Alc{\'a}zar}, {{\"O}zel},
  {Dav{\'e}}, {Katz}, {Kollmeier}  \& {Oppenheimer}}{{Angl{\'e}s-Alc{\'a}zar}
  et~al.}{2015}]{Angles-Alcazar2015}
{Angl{\'e}s-Alc{\'a}zar} D.,  {{\"O}zel} F.,  {Dav{\'e}} R.,  {Katz} N.,
  {Kollmeier} J.~A.,   {Oppenheimer} B.~D.,  2015, \mn@doi [\apj]
  {10.1088/0004-637X/800/2/127}, \href
  {https://ui.adsabs.harvard.edu/abs/2015ApJ...800..127A} {800, 127}

\bibitem[\protect\citeauthoryear{{Angl{\'e}s-Alc{\'a}zar}, {Dav{\'e}},
  {Faucher-Gigu{\`e}re}, {{\"O}zel}  \& {Hopkins}}{{Angl{\'e}s-Alc{\'a}zar}
  et~al.}{2017a}]{Angles-Alcazar2017a}
{Angl{\'e}s-Alc{\'a}zar} D.,  {Dav{\'e}} R.,  {Faucher-Gigu{\`e}re} C.-A.,
  {{\"O}zel} F.,   {Hopkins} P.~F.,  2017a, \mn@doi [\mnras]
  {10.1093/mnras/stw2565}, \href
  {https://ui.adsabs.harvard.edu/abs/2017MNRAS.464.2840A} {464, 2840}

\bibitem[\protect\citeauthoryear{{Angl{\'e}s-Alc{\'a}zar},
  {Faucher-Gigu{\`e}re}, {Kere{\v{s}}}, {Hopkins}, {Quataert}  \&
  {Murray}}{{Angl{\'e}s-Alc{\'a}zar} et~al.}{2017b}]{Angles-Alcazar2017b}
{Angl{\'e}s-Alc{\'a}zar} D.,  {Faucher-Gigu{\`e}re} C.-A.,  {Kere{\v{s}}} D.,
  {Hopkins} P.~F.,  {Quataert} E.,   {Murray} N.,  2017b, \mn@doi [\mnras]
  {10.1093/mnras/stx1517}, \href
  {https://ui.adsabs.harvard.edu/abs/2017MNRAS.470.4698A} {470, 4698}

\bibitem[\protect\citeauthoryear{{Angl{\'e}s-Alc{\'a}zar},
  {Faucher-Gigu{\`e}re}, {Quataert}, {Hopkins}, {Feldmann}, {Torrey}, {Wetzel}
  \& {Kere{\v{s}}}}{{Angl{\'e}s-Alc{\'a}zar}
  et~al.}{2017c}]{Angles-Alcazar2017c}
{Angl{\'e}s-Alc{\'a}zar} D.,  {Faucher-Gigu{\`e}re} C.-A.,  {Quataert} E.,
  {Hopkins} P.~F.,  {Feldmann} R.,  {Torrey} P.,  {Wetzel} A.,   {Kere{\v{s}}}
  D.,  2017c, \mn@doi [\mnras] {10.1093/mnrasl/slx161}, \href
  {https://ui.adsabs.harvard.edu/abs/2017MNRAS.472L.109A} {472, L109}

\bibitem[\protect\citeauthoryear{{Angl{\'e}s-Alc{\'a}zar}
  et~al.,}{{Angl{\'e}s-Alc{\'a}zar} et~al.}{2021}]{Angles-Alcazar2021}
{Angl{\'e}s-Alc{\'a}zar} D.,  et~al., 2021, \mn@doi [\apj]
  {10.3847/1538-4357/ac09e8}, \href
  {https://ui.adsabs.harvard.edu/abs/2021ApJ...917...53A} {917, 53}

\bibitem[\protect\citeauthoryear{{Bah{\'e}} et~al.,}{{Bah{\'e}}
  et~al.}{2022}]{Bahe2022}
{Bah{\'e}} Y.~M.,  et~al., 2022, \mn@doi [\mnras] {10.1093/mnras/stac1339},
  \href {https://ui.adsabs.harvard.edu/abs/2022MNRAS.516..167B} {516, 167}

\bibitem[\protect\citeauthoryear{{Baldassare}, {Dickey}, {Geha}  \&
  {Reines}}{{Baldassare} et~al.}{2020}]{Baldassare2020}
{Baldassare} V.~F.,  {Dickey} C.,  {Geha} M.,   {Reines} A.~E.,  2020, \mn@doi
  [\apjl] {10.3847/2041-8213/aba0c1}, \href
  {https://ui.adsabs.harvard.edu/abs/2020ApJ...898L...3B} {898, L3}

\bibitem[\protect\citeauthoryear{{Benedetti} et~al.,}{{Benedetti}
  et~al.}{2023}]{Benedetti2023}
{Benedetti} J. P.~V.,  et~al., 2023, \mn@doi [\mnras] {10.1093/mnras/stad1148},
  \href {https://ui.adsabs.harvard.edu/abs/2023MNRAS.522.2570B} {522, 2570}

\bibitem[\protect\citeauthoryear{{Blecha}, {Snyder}, {Satyapal}  \&
  {Ellison}}{{Blecha} et~al.}{2018}]{Blecha2018}
{Blecha} L.,  {Snyder} G.~F.,  {Satyapal} S.,   {Ellison} S.~L.,  2018, \mn@doi
  [\mnras] {10.1093/mnras/sty1274}, \href
  {https://ui.adsabs.harvard.edu/abs/2018MNRAS.478.3056B} {478, 3056}

\bibitem[\protect\citeauthoryear{{Blumenthal} \& {Barnes}}{{Blumenthal} \&
  {Barnes}}{2018}]{Blumenthal2018}
{Blumenthal} K.~A.,  {Barnes} J.~E.,  2018, \mn@doi [\mnras]
  {10.1093/mnras/sty1605}, \href
  {https://ui.adsabs.harvard.edu/abs/2018MNRAS.479.3952B} {479, 3952}

\bibitem[\protect\citeauthoryear{{B{\"o}ker} et~al.,}{{B{\"o}ker}
  et~al.}{2022}]{Boker2022}
{B{\"o}ker} T.,  et~al., 2022, \mn@doi [\aap] {10.1051/0004-6361/202142589},
  \href {https://ui.adsabs.harvard.edu/abs/2022A&A...661A..82B} {661, A82}

\bibitem[\protect\citeauthoryear{{Borrow}, {Angl{\'e}s-Alc{\'a}zar}  \&
  {Dav{\'e}}}{{Borrow} et~al.}{2020}]{Borrow2020}
{Borrow} J.,  {Angl{\'e}s-Alc{\'a}zar} D.,   {Dav{\'e}} R.,  2020, \mn@doi
  [\mnras] {10.1093/mnras/stz3428}, \href
  {https://ui.adsabs.harvard.edu/abs/2020MNRAS.491.6102B} {491, 6102}

\bibitem[\protect\citeauthoryear{{Bower}, {Schaye}, {Frenk}, {Theuns},
  {Schaller}, {Crain}  \& {McAlpine}}{{Bower} et~al.}{2017}]{Bower2017}
{Bower} R.~G.,  {Schaye} J.,  {Frenk} C.~S.,  {Theuns} T.,  {Schaller} M.,
  {Crain} R.~A.,   {McAlpine} S.,  2017, \mn@doi [\mnras]
  {10.1093/mnras/stw2735}, \href
  {https://ui.adsabs.harvard.edu/abs/2017MNRAS.465...32B} {465, 32}

\bibitem[\protect\citeauthoryear{{Boyett} et~al.,}{{Boyett}
  et~al.}{2024}]{Boyett2024}
{Boyett} K.,  et~al., 2024, \mn@doi [\mnras] {10.1093/mnras/stae2430}, \href
  {https://ui.adsabs.harvard.edu/abs/2024MNRAS.535.1796B} {535, 1796}

\bibitem[\protect\citeauthoryear{{Byrne-Mamahit}, {Ellison}, {Patton},
  {Wilkinson}, {Ferreira}  \& {Bottrell}}{{Byrne-Mamahit}
  et~al.}{2025}]{Byrne-Mamahit2025}
{Byrne-Mamahit} S.,  {Ellison} S.~L.,  {Patton} D.~R.,  {Wilkinson} S.,
  {Ferreira} L.,   {Bottrell} C.,  2025, \mn@doi [arXiv e-prints]
  {10.48550/arXiv.2510.12738}, \href
  {https://ui.adsabs.harvard.edu/abs/2025arXiv251012738B} {p. arXiv:2510.12738}

\bibitem[\protect\citeauthoryear{{Byrne}, {Faucher-Gigu{\`e}re}, {Stern},
  {Angl{\'e}s-Alc{\'a}zar}, {Wellons}, {Gurvich}  \& {Hopkins}}{{Byrne}
  et~al.}{2023}]{Byrne2023}
{Byrne} L.,  {Faucher-Gigu{\`e}re} C.-A.,  {Stern} J.,
  {Angl{\'e}s-Alc{\'a}zar} D.,  {Wellons} S.,  {Gurvich} A.~B.,   {Hopkins}
  P.~F.,  2023, \mn@doi [\mnras] {10.1093/mnras/stad171}, \href
  {https://ui.adsabs.harvard.edu/abs/2023MNRAS.520..722B} {520, 722}

\bibitem[\protect\citeauthoryear{{Byrne} et~al.,}{{Byrne}
  et~al.}{2024}]{Byrne2024}
{Byrne} L.,  et~al., 2024, \mn@doi [\apj] {10.3847/1538-4357/ad67ca}, \href
  {https://ui.adsabs.harvard.edu/abs/2024ApJ...973..149B} {973, 149}

\bibitem[\protect\citeauthoryear{{Canalizo} \& {Stockton}}{{Canalizo} \&
  {Stockton}}{2001}]{Canalizo2001}
{Canalizo} G.,  {Stockton} A.,  2001, \mn@doi [\apj] {10.1086/321520}, \href
  {https://ui.adsabs.harvard.edu/abs/2001ApJ...555..719C} {555, 719}

\bibitem[\protect\citeauthoryear{{Carnall} et~al.,}{{Carnall}
  et~al.}{2024}]{Carnall2024}
{Carnall} A.~C.,  et~al., 2024, \mn@doi [\mnras] {10.1093/mnras/stae2092},
  \href {https://ui.adsabs.harvard.edu/abs/2024MNRAS.534..325C} {534, 325}

\bibitem[\protect\citeauthoryear{{Chastenet} et~al.,}{{Chastenet}
  et~al.}{2024}]{Chastenet2024}
{Chastenet} J.,  et~al., 2024, \mn@doi [\aap] {10.1051/0004-6361/202451033},
  \href {https://ui.adsabs.harvard.edu/abs/2024A&A...690A.348C} {690, A348}

\bibitem[\protect\citeauthoryear{{Chisholm}, {Tremonti}, {Leitherer}  \&
  {Chen}}{{Chisholm} et~al.}{2017}]{Chisholm2017}
{Chisholm} J.,  {Tremonti} C.~A.,  {Leitherer} C.,   {Chen} Y.,  2017, \mn@doi
  [\mnras] {10.1093/mnras/stx1164}, \href
  {https://ui.adsabs.harvard.edu/abs/2017MNRAS.469.4831C} {469, 4831}

\bibitem[\protect\citeauthoryear{{Choi}, {Ostriker}, {Naab}, {Somerville},
  {Hirschmann}, {N{\'u}{\~n}ez}, {Hu}  \& {Oser}}{{Choi}
  et~al.}{2017}]{Choi2017}
{Choi} E.,  {Ostriker} J.~P.,  {Naab} T.,  {Somerville} R.~S.,  {Hirschmann}
  M.,  {N{\'u}{\~n}ez} A.,  {Hu} C.-Y.,   {Oser} L.,  2017, \mn@doi [\apj]
  {10.3847/1538-4357/aa7849}, \href
  {https://ui.adsabs.harvard.edu/abs/2017ApJ...844...31C} {844, 31}

\bibitem[\protect\citeauthoryear{{Choi}, {Somerville}, {Ostriker}, {Naab}  \&
  {Hirschmann}}{{Choi} et~al.}{2018}]{Choi2018}
{Choi} E.,  {Somerville} R.~S.,  {Ostriker} J.~P.,  {Naab} T.,   {Hirschmann}
  M.,  2018, \mn@doi [\apj] {10.3847/1538-4357/aae076}, \href
  {https://ui.adsabs.harvard.edu/abs/2018ApJ...866...91C} {866, 91}

\bibitem[\protect\citeauthoryear{{Choi}, {Somerville}, {Ostriker}, {Hirschmann}
   \& {Naab}}{{Choi} et~al.}{2024}]{Choi2024}
{Choi} E.,  {Somerville} R.~S.,  {Ostriker} J.~P.,  {Hirschmann} M.,   {Naab}
  T.,  2024, \mn@doi [\apj] {10.3847/1538-4357/ad245a}, \href
  {https://ui.adsabs.harvard.edu/abs/2024ApJ...964...54C} {964, 54}

\bibitem[\protect\citeauthoryear{{Christensen}, {Dav{\'e}}, {Governato},
  {Pontzen}, {Brooks}, {Munshi}, {Quinn}  \& {Wadsley}}{{Christensen}
  et~al.}{2016}]{Christensen2016}
{Christensen} C.~R.,  {Dav{\'e}} R.,  {Governato} F.,  {Pontzen} A.,  {Brooks}
  A.,  {Munshi} F.,  {Quinn} T.,   {Wadsley} J.,  2016, \mn@doi [\apj]
  {10.3847/0004-637X/824/1/57}, \href
  {https://ui.adsabs.harvard.edu/abs/2016ApJ...824...57C} {824, 57}

\bibitem[\protect\citeauthoryear{{Ciotti} \& {Ostriker}}{{Ciotti} \&
  {Ostriker}}{2007}]{Ciotti2007}
{Ciotti} L.,  {Ostriker} J.~P.,  2007, \mn@doi [\apj] {10.1086/519833}, \href
  {https://ui.adsabs.harvard.edu/abs/2007ApJ...665.1038C} {665, 1038}

\bibitem[\protect\citeauthoryear{{Cisternas} et~al.,}{{Cisternas}
  et~al.}{2011}]{Cisternas2011}
{Cisternas} M.,  et~al., 2011, \mn@doi [\apj] {10.1088/0004-637X/726/2/57},
  \href {https://ui.adsabs.harvard.edu/abs/2011ApJ...726...57C} {726, 57}

\bibitem[\protect\citeauthoryear{{Cochrane} et~al.,}{{Cochrane}
  et~al.}{2023}]{Cochrane2023a}
{Cochrane} R.~K.,  et~al., 2023, \mn@doi [\mnras] {10.1093/mnras/stad1528},
  \href {https://ui.adsabs.harvard.edu/abs/2023MNRAS.523.2409C} {523, 2409}

\bibitem[\protect\citeauthoryear{{Cole} et~al.,}{{Cole}
  et~al.}{2025}]{Cole2025}
{Cole} J.~W.,  et~al., 2025, \mn@doi [\apj] {10.3847/1538-4357/ad9a6a}, \href
  {https://ui.adsabs.harvard.edu/abs/2025ApJ...979..193C} {979, 193}

\bibitem[\protect\citeauthoryear{{Combes}}{{Combes}}{2023}]{Combes2023}
{Combes} F.,  2023, \mn@doi [Galaxies] {10.3390/galaxies11060120}, \href
  {https://ui.adsabs.harvard.edu/abs/2023Galax..11..120C} {11, 120}

\bibitem[\protect\citeauthoryear{{Costa}, {Rosdahl}, {Sijacki}  \&
  {Haehnelt}}{{Costa} et~al.}{2018}]{Costa2018b}
{Costa} T.,  {Rosdahl} J.,  {Sijacki} D.,   {Haehnelt} M.~G.,  2018, \mn@doi
  [\mnras] {10.1093/mnras/sty1514}, \href
  {https://ui.adsabs.harvard.edu/abs/2018MNRAS.479.2079C} {479, 2079}

\bibitem[\protect\citeauthoryear{{Cotini}, {Ripamonti}, {Caccianiga}, {Colpi},
  {Della Ceca}, {Mapelli}, {Severgnini}  \& {Segreto}}{{Cotini}
  et~al.}{2013}]{Cotini2013}
{Cotini} S.,  {Ripamonti} E.,  {Caccianiga} A.,  {Colpi} M.,  {Della Ceca} R.,
  {Mapelli} M.,  {Severgnini} P.,   {Segreto} A.,  2013, \mn@doi [\mnras]
  {10.1093/mnras/stt358}, \href
  {https://ui.adsabs.harvard.edu/abs/2013MNRAS.431.2661C} {431, 2661}

\bibitem[\protect\citeauthoryear{{Covelo-Paz} et~al.,}{{Covelo-Paz}
  et~al.}{2025}]{Covelo-Paz2025}
{Covelo-Paz} A.,  et~al., 2025, \mn@doi [arXiv e-prints]
  {10.48550/arXiv.2506.22540}, \href
  {https://ui.adsabs.harvard.edu/abs/2025arXiv250622540C} {p. arXiv:2506.22540}

\bibitem[\protect\citeauthoryear{{Dav{\'e}}, {Angl{\'e}s-Alc{\'a}zar},
  {Narayanan}, {Li}, {Rafieferantsoa}  \& {Appleby}}{{Dav{\'e}}
  et~al.}{2019}]{Dave2019}
{Dav{\'e}} R.,  {Angl{\'e}s-Alc{\'a}zar} D.,  {Narayanan} D.,  {Li} Q.,
  {Rafieferantsoa} M.~H.,   {Appleby} S.,  2019, \mn@doi [\mnras]
  {10.1093/mnras/stz937}, \href
  {https://ui.adsabs.harvard.edu/abs/2019MNRAS.486.2827D} {486, 2827}

\bibitem[\protect\citeauthoryear{{Davis}, {Graham}  \& {Cameron}}{{Davis}
  et~al.}{2018}]{Davis2018}
{Davis} B.~L.,  {Graham} A.~W.,   {Cameron} E.,  2018, \mn@doi [\apj]
  {10.3847/1538-4357/aae820}, \href
  {https://ui.adsabs.harvard.edu/abs/2018ApJ...869..113D} {869, 113}

\bibitem[\protect\citeauthoryear{{Davis}, {Graham}  \& {Cameron}}{{Davis}
  et~al.}{2019}]{Davis2019}
{Davis} B.~L.,  {Graham} A.~W.,   {Cameron} E.,  2019, \mn@doi [\apj]
  {10.3847/1538-4357/aaf3b8}, \href
  {https://ui.adsabs.harvard.edu/abs/2019ApJ...873...85D} {873, 85}

\bibitem[\protect\citeauthoryear{{Dehnen} \& {King}}{{Dehnen} \&
  {King}}{2013}]{Dehnen2013}
{Dehnen} W.,  {King} A.,  2013, \mn@doi [\apjl] {10.1088/2041-8205/777/2/L28},
  \href {https://ui.adsabs.harvard.edu/abs/2013ApJ...777L..28D} {777, L28}

\bibitem[\protect\citeauthoryear{{D{\'\i}az-Santos} et~al.,}{{D{\'\i}az-Santos}
  et~al.}{2018}]{Diaz-Santos2018}
{D{\'\i}az-Santos} T.,  et~al., 2018, \mn@doi [Science]
  {10.1126/science.aap7605}, \href
  {https://ui.adsabs.harvard.edu/abs/2018Sci...362.1034D} {362, 1034}

\bibitem[\protect\citeauthoryear{{Dubois}, {Volonteri}, {Silk}, {Devriendt},
  {Slyz}  \& {Teyssier}}{{Dubois} et~al.}{2015}]{Dubois2015}
{Dubois} Y.,  {Volonteri} M.,  {Silk} J.,  {Devriendt} J.,  {Slyz} A.,
  {Teyssier} R.,  2015, \mn@doi [\mnras] {10.1093/mnras/stv1416}, \href
  {https://ui.adsabs.harvard.edu/abs/2015MNRAS.452.1502D} {452, 1502}

\bibitem[\protect\citeauthoryear{{Dubois} et~al.,}{{Dubois}
  et~al.}{2021}]{Dubois2021}
{Dubois} Y.,  et~al., 2021, \mn@doi [\aap] {10.1051/0004-6361/202039429}, \href
  {https://ui.adsabs.harvard.edu/abs/2021A&A...651A.109D} {651, A109}

\bibitem[\protect\citeauthoryear{{Dullo}, {Bouquin}, {Gil de Paz}, {Knapen}  \&
  {Gorgas}}{{Dullo} et~al.}{2020}]{Dullo2020}
{Dullo} B.~T.,  {Bouquin} A. Y.~K.,  {Gil de Paz} A.,  {Knapen} J.~H.,
  {Gorgas} J.,  2020, \mn@doi [\apj] {10.3847/1538-4357/ab9dff}, \href
  {https://ui.adsabs.harvard.edu/abs/2020ApJ...898...83D} {898, 83}

\bibitem[\protect\citeauthoryear{{Egorov} et~al.,}{{Egorov}
  et~al.}{2023}]{Egorov2023}
{Egorov} O.~V.,  et~al., 2023, \mn@doi [\aap] {10.1051/0004-6361/202346919},
  \href {https://ui.adsabs.harvard.edu/abs/2023A&A...678A.153E} {678, A153}

\bibitem[\protect\citeauthoryear{{Eilers}, {Davies}, {Hennawi}, {Prochaska},
  {Luki{\'c}}  \& {Mazzucchelli}}{{Eilers} et~al.}{2017}]{Eilers2017}
{Eilers} A.-C.,  {Davies} F.~B.,  {Hennawi} J.~F.,  {Prochaska} J.~X.,
  {Luki{\'c}} Z.,   {Mazzucchelli} C.,  2017, \mn@doi [\apj]
  {10.3847/1538-4357/aa6c60}, \href
  {https://ui.adsabs.harvard.edu/abs/2017ApJ...840...24E} {840, 24}

\bibitem[\protect\citeauthoryear{{Eilers} et~al.,}{{Eilers}
  et~al.}{2025}]{Eilers2025}
{Eilers} A.-C.,  et~al., 2025, \mn@doi [\apjl] {10.3847/2041-8213/ae057a},
  \href {https://ui.adsabs.harvard.edu/abs/2025ApJ...991L..40E} {991, L40}

\bibitem[\protect\citeauthoryear{{Ellison}, {Viswanathan}, {Patton},
  {Bottrell}, {McConnachie}, {Gwyn}  \& {Cuillandre}}{{Ellison}
  et~al.}{2019}]{Ellison2019}
{Ellison} S.~L.,  {Viswanathan} A.,  {Patton} D.~R.,  {Bottrell} C.,
  {McConnachie} A.~W.,  {Gwyn} S.,   {Cuillandre} J.-C.,  2019, \mn@doi
  [\mnras] {10.1093/mnras/stz1431}, \href
  {https://ui.adsabs.harvard.edu/abs/2019MNRAS.487.2491E} {487, 2491}

\bibitem[\protect\citeauthoryear{{Emami}, {Siana}, {Weisz}, {Johnson}, {Ma}  \&
  {El-Badry}}{{Emami} et~al.}{2019}]{Emami2019}
{Emami} N.,  {Siana} B.,  {Weisz} D.~R.,  {Johnson} B.~D.,  {Ma} X.,
  {El-Badry} K.,  2019, \mn@doi [\apj] {10.3847/1538-4357/ab211a}, \href
  {https://ui.adsabs.harvard.edu/abs/2019ApJ...881...71E} {881, 71}

\bibitem[\protect\citeauthoryear{{Endsley}, {Chisholm}, {Stark}, {Topping}  \&
  {Whitler}}{{Endsley} et~al.}{2025}]{Endsley2025}
{Endsley} R.,  {Chisholm} J.,  {Stark} D.~P.,  {Topping} M.~W.,   {Whitler} L.,
   2025, \mn@doi [\apj] {10.3847/1538-4357/addc74}, \href
  {https://ui.adsabs.harvard.edu/abs/2025ApJ...987..189E} {987, 189}

\bibitem[\protect\citeauthoryear{{Engel} et~al.,}{{Engel}
  et~al.}{2010}]{Engel2010}
{Engel} H.,  et~al., 2010, \mn@doi [\apj] {10.1088/0004-637X/724/1/233}, \href
  {https://ui.adsabs.harvard.edu/abs/2010ApJ...724..233E} {724, 233}

\bibitem[\protect\citeauthoryear{{Faisst}, {Capak}, {Emami}, {Tacchella}  \&
  {Larson}}{{Faisst} et~al.}{2019}]{Faisst2019}
{Faisst} A.~L.,  {Capak} P.~L.,  {Emami} N.,  {Tacchella} S.,   {Larson} K.~L.,
   2019, \mn@doi [\apj] {10.3847/1538-4357/ab425b}, \href
  {https://ui.adsabs.harvard.edu/abs/2019ApJ...884..133F} {884, 133}

\bibitem[\protect\citeauthoryear{{Faucher-Gigu{\`e}re}}{{Faucher-Gigu{\`e}re}}{2018}]{Faucher-Giguere2018}
{Faucher-Gigu{\`e}re} C.-A.,  2018, \mn@doi [\mnras] {10.1093/mnras/stx2595},
  \href {https://ui.adsabs.harvard.edu/abs/2018MNRAS.473.3717F} {473, 3717}

\bibitem[\protect\citeauthoryear{{Faucher-Gigu{\`e}re} \&
  {Oh}}{{Faucher-Gigu{\`e}re} \& {Oh}}{2023}]{Faucher-Giguere2023}
{Faucher-Gigu{\`e}re} C.-A.,  {Oh} S.~P.,  2023, \mn@doi [\araa]
  {10.1146/annurev-astro-052920-125203}, \href
  {https://ui.adsabs.harvard.edu/abs/2023ARA&A..61..131F} {61, 131}

\bibitem[\protect\citeauthoryear{{Feldmann}, {Hopkins}, {Quataert},
  {Faucher-Gigu{\`e}re}  \& {Kere{\v{s}}}}{{Feldmann}
  et~al.}{2016}]{Feldmann2016}
{Feldmann} R.,  {Hopkins} P.~F.,  {Quataert} E.,  {Faucher-Gigu{\`e}re} C.-A.,
   {Kere{\v{s}}} D.,  2016, \mn@doi [\mnras] {10.1093/mnrasl/slw014}, \href
  {https://ui.adsabs.harvard.edu/abs/2016MNRAS.458L..14F} {458, L14}

\bibitem[\protect\citeauthoryear{{Feldmann}, {Quataert}, {Hopkins},
  {Faucher-Gigu{\`e}re}  \& {Kere{\v{s}}}}{{Feldmann}
  et~al.}{2017}]{Feldmann2017}
{Feldmann} R.,  {Quataert} E.,  {Hopkins} P.~F.,  {Faucher-Gigu{\`e}re} C.-A.,
   {Kere{\v{s}}} D.,  2017, \mn@doi [\mnras] {10.1093/mnras/stx1120}, \href
  {https://ui.adsabs.harvard.edu/abs/2017MNRAS.470.1050F} {470, 1050}

\bibitem[\protect\citeauthoryear{{Gandhi} et~al.,}{{Gandhi}
  et~al.}{2024}]{Gandhi2024}
{Gandhi} P.~J.,  et~al., 2024, \mn@doi [\mnras] {10.1093/mnras/stae1584}, \href
  {https://ui.adsabs.harvard.edu/abs/2024MNRAS.533.1059G} {533, 1059}

\bibitem[\protect\citeauthoryear{{Garc{\'\i}a-Burillo}, {Combes}, {Schinnerer},
  {Boone}  \& {Hunt}}{{Garc{\'\i}a-Burillo} et~al.}{2005}]{Garcia-Burillo2005}
{Garc{\'\i}a-Burillo} S.,  {Combes} F.,  {Schinnerer} E.,  {Boone} F.,   {Hunt}
  L.~K.,  2005, \mn@doi [\aap] {10.1051/0004-6361:20052900}, \href
  {https://ui.adsabs.harvard.edu/abs/2005A&A...441.1011G} {441, 1011}

\bibitem[\protect\citeauthoryear{{Gebhardt} et~al.,}{{Gebhardt}
  et~al.}{2024}]{Gebhardt2024}
{Gebhardt} M.,  et~al., 2024, \mn@doi [\mnras] {10.1093/mnras/stae817}, \href
  {https://ui.adsabs.harvard.edu/abs/2024MNRAS.529.4896G} {529, 4896}

\bibitem[\protect\citeauthoryear{{Graham} \& {Scott}}{{Graham} \&
  {Scott}}{2013}]{Graham2013}
{Graham} A.~W.,  {Scott} N.,  2013, \mn@doi [\apj]
  {10.1088/0004-637X/764/2/151}, \href
  {https://ui.adsabs.harvard.edu/abs/2013ApJ...764..151G} {764, 151}

\bibitem[\protect\citeauthoryear{{Grand} et~al.,}{{Grand}
  et~al.}{2019}]{Grand2019}
{Grand} R. J.~J.,  et~al., 2019, \mn@doi [\mnras] {10.1093/mnras/stz2928},
  \href {https://ui.adsabs.harvard.edu/abs/2019MNRAS.490.4786G} {490, 4786}

\bibitem[\protect\citeauthoryear{{Gurvich} et~al.,}{{Gurvich}
  et~al.}{2023}]{Gurvich2023}
{Gurvich} A.~B.,  et~al., 2023, \mn@doi [\mnras] {10.1093/mnras/stac3712},
  \href {https://ui.adsabs.harvard.edu/abs/2023MNRAS.519.2598G} {519, 2598}

\bibitem[\protect\citeauthoryear{{Habouzit}, {Volonteri}  \&
  {Dubois}}{{Habouzit} et~al.}{2017}]{Habouzit2017}
{Habouzit} M.,  {Volonteri} M.,   {Dubois} Y.,  2017, \mn@doi [\mnras]
  {10.1093/mnras/stx666}, \href
  {https://ui.adsabs.harvard.edu/abs/2017MNRAS.468.3935H} {468, 3935}

\bibitem[\protect\citeauthoryear{{Habouzit} et~al.,}{{Habouzit}
  et~al.}{2021}]{Habouzit2021}
{Habouzit} M.,  et~al., 2021, \mn@doi [\mnras] {10.1093/mnras/stab496}, \href
  {https://ui.adsabs.harvard.edu/abs/2021MNRAS.503.1940H} {503, 1940}

\bibitem[\protect\citeauthoryear{{Hafen} et~al.,}{{Hafen}
  et~al.}{2019}]{Hafen2019}
{Hafen} Z.,  et~al., 2019, \mn@doi [\mnras] {10.1093/mnras/stz1773}, \href
  {https://ui.adsabs.harvard.edu/abs/2019MNRAS.488.1248H} {488, 1248}

\bibitem[\protect\citeauthoryear{{Hafen} et~al.,}{{Hafen}
  et~al.}{2020}]{Hafen2020}
{Hafen} Z.,  et~al., 2020, \mn@doi [\mnras] {10.1093/mnras/staa902}, \href
  {https://ui.adsabs.harvard.edu/abs/2020MNRAS.494.3581H} {494, 3581}

\bibitem[\protect\citeauthoryear{{Hafen} et~al.,}{{Hafen}
  et~al.}{2022}]{Hafen2022}
{Hafen} Z.,  et~al., 2022, \mn@doi [\mnras] {10.1093/mnras/stac1603}, \href
  {https://ui.adsabs.harvard.edu/abs/2022MNRAS.514.5056H} {514, 5056}

\bibitem[\protect\citeauthoryear{{Harikane} et~al.,}{{Harikane}
  et~al.}{2023}]{Harikane2023}
{Harikane} Y.,  et~al., 2023, \mn@doi [\apjs] {10.3847/1538-4365/acaaa9}, \href
  {https://ui.adsabs.harvard.edu/abs/2023ApJS..265....5H} {265, 5}

\bibitem[\protect\citeauthoryear{{Hayward} \& {Hopkins}}{{Hayward} \&
  {Hopkins}}{2017}]{Hayward2017}
{Hayward} C.~C.,  {Hopkins} P.~F.,  2017, \mn@doi [\mnras]
  {10.1093/mnras/stw2888}, \href
  {https://ui.adsabs.harvard.edu/abs/2017MNRAS.465.1682H} {465, 1682}

\bibitem[\protect\citeauthoryear{{Heckman}, {Borthakur}, {Wild}, {Schiminovich}
   \& {Bordoloi}}{{Heckman} et~al.}{2017}]{Heckman2017}
{Heckman} T.,  {Borthakur} S.,  {Wild} V.,  {Schiminovich} D.,   {Bordoloi} R.,
   2017, \mn@doi [\apj] {10.3847/1538-4357/aa80dc}, \href
  {https://ui.adsabs.harvard.edu/abs/2017ApJ...846..151H} {846, 151}

\bibitem[\protect\citeauthoryear{{Hinshaw} et~al.,}{{Hinshaw}
  et~al.}{2013}]{Hinshaw2013}
{Hinshaw} G.,  et~al., 2013, \mn@doi [\apjs] {10.1088/0067-0049/208/2/19},
  \href {https://ui.adsabs.harvard.edu/abs/2013ApJS..208...19H} {208, 19}

\bibitem[\protect\citeauthoryear{{Hlavacek-Larrondo}
  et~al.,}{{Hlavacek-Larrondo} et~al.}{2015}]{Hlavacek-Larrondo2015}
{Hlavacek-Larrondo} J.,  et~al., 2015, \mn@doi [\apj]
  {10.1088/0004-637X/805/1/35}, \href
  {https://ui.adsabs.harvard.edu/abs/2015ApJ...805...35H} {805, 35}

\bibitem[\protect\citeauthoryear{{Ho}, {Martin}  \& {Turner}}{{Ho}
  et~al.}{2019}]{Ho2019}
{Ho} S.~H.,  {Martin} C.~L.,   {Turner} M.~L.,  2019, \mn@doi [\apj]
  {10.3847/1538-4357/ab0ec2}, \href
  {https://ui.adsabs.harvard.edu/abs/2019ApJ...875...54H} {875, 54}

\bibitem[\protect\citeauthoryear{{Hopkins}}{{Hopkins}}{2015}]{Hopkins2015gizmo}
{Hopkins} P.~F.,  2015, \mn@doi [\mnras] {10.1093/mnras/stv195}, \href
  {https://ui.adsabs.harvard.edu/abs/2015MNRAS.450...53H} {450, 53}

\bibitem[\protect\citeauthoryear{{Hopkins} \& {Quataert}}{{Hopkins} \&
  {Quataert}}{2010}]{Hopkins&Quataert2010}
{Hopkins} P.~F.,  {Quataert} E.,  2010, \mn@doi [\mnras]
  {10.1111/j.1365-2966.2010.17064.x}, \href
  {https://ui.adsabs.harvard.edu/abs/2010MNRAS.407.1529H} {407, 1529}

\bibitem[\protect\citeauthoryear{{Hopkins} \& {Quataert}}{{Hopkins} \&
  {Quataert}}{2011}]{Hopkins&Quataert2011}
{Hopkins} P.~F.,  {Quataert} E.,  2011, \mn@doi [\mnras]
  {10.1111/j.1365-2966.2011.18542.x}, \href
  {https://ui.adsabs.harvard.edu/abs/2011MNRAS.415.1027H} {415, 1027}

\bibitem[\protect\citeauthoryear{{Hopkins}, {Hernquist}, {Cox}, {Robertson}  \&
  {Springel}}{{Hopkins} et~al.}{2006}]{Hopkins2006}
{Hopkins} P.~F.,  {Hernquist} L.,  {Cox} T.~J.,  {Robertson} B.,   {Springel}
  V.,  2006, \mn@doi [\apjs] {10.1086/499493}, \href
  {https://ui.adsabs.harvard.edu/abs/2006ApJS..163...50H} {163, 50}

\bibitem[\protect\citeauthoryear{{Hopkins}, {Kere{\v{s}}}, {O{\~n}orbe},
  {Faucher-Gigu{\`e}re}, {Quataert}, {Murray}  \& {Bullock}}{{Hopkins}
  et~al.}{2014}]{Hopkins2014}
{Hopkins} P.~F.,  {Kere{\v{s}}} D.,  {O{\~n}orbe} J.,  {Faucher-Gigu{\`e}re}
  C.-A.,  {Quataert} E.,  {Murray} N.,   {Bullock} J.~S.,  2014, \mn@doi
  [\mnras] {10.1093/mnras/stu1738}, \href
  {https://ui.adsabs.harvard.edu/abs/2014MNRAS.445..581H} {445, 581}

\bibitem[\protect\citeauthoryear{{Hopkins} et~al.,}{{Hopkins}
  et~al.}{2018}]{Hopkins2018}
{Hopkins} P.~F.,  et~al., 2018, \mn@doi [\mnras] {10.1093/mnras/sty1690}, \href
  {https://ui.adsabs.harvard.edu/abs/2018MNRAS.480..800H} {480, 800}

\bibitem[\protect\citeauthoryear{{Hopkins} et~al.,}{{Hopkins}
  et~al.}{2023a}]{Hopkins2023_fire3}
{Hopkins} P.~F.,  et~al., 2023a, \mn@doi [\mnras] {10.1093/mnras/stac3489},
  \href {https://ui.adsabs.harvard.edu/abs/2023MNRAS.519.3154H} {519, 3154}

\bibitem[\protect\citeauthoryear{{Hopkins} et~al.,}{{Hopkins}
  et~al.}{2023b}]{Hopkins2023}
{Hopkins} P.~F.,  et~al., 2023b, \mn@doi [\mnras] {10.1093/mnras/stad1902},
  \href {https://ui.adsabs.harvard.edu/abs/2023MNRAS.525.2241H} {525, 2241}

\bibitem[\protect\citeauthoryear{{Hopkins} et~al.,}{{Hopkins}
  et~al.}{2024}]{Hopkins2024_zoom1}
{Hopkins} P.~F.,  et~al., 2024, \mn@doi [The Open Journal of Astrophysics]
  {10.21105/astro.2309.13115}, \href
  {https://ui.adsabs.harvard.edu/abs/2024OJAp....7E..18H} {7, 18}

\bibitem[\protect\citeauthoryear{{Jogee}}{{Jogee}}{2006}]{Jogee2006}
{Jogee} S.,  2006, in {Alloin} D.,  ed., , Vol.~693, Physics of Active Galactic
  Nuclei at all Scales.
p.~143, \mn@doi{10.1007/3-540-34621-X_6}

\bibitem[\protect\citeauthoryear{{Kaviraj}}{{Kaviraj}}{2014}]{Kaviraj2014}
{Kaviraj} S.,  2014, \mn@doi [\mnras] {10.1093/mnras/stu338}, \href
  {https://ui.adsabs.harvard.edu/abs/2014MNRAS.440.2944K} {440, 2944}

\bibitem[\protect\citeauthoryear{{Kim} et~al.,}{{Kim} et~al.}{2020}]{Kim2020}
{Kim} C.-G.,  et~al., 2020, \mn@doi [\apj] {10.3847/1538-4357/aba962}, \href
  {https://ui.adsabs.harvard.edu/abs/2020ApJ...900...61K} {900, 61}

\bibitem[\protect\citeauthoryear{{Kocevski} et~al.,}{{Kocevski}
  et~al.}{2012}]{Kocevski2012}
{Kocevski} D.~D.,  et~al., 2012, \mn@doi [\apj] {10.1088/0004-637X/744/2/148},
  \href {https://ui.adsabs.harvard.edu/abs/2012ApJ...744..148K} {744, 148}

\bibitem[\protect\citeauthoryear{{Kocevski} et~al.,}{{Kocevski}
  et~al.}{2023}]{Kocevski2023}
{Kocevski} D.~D.,  et~al., 2023, \mn@doi [\apjl] {10.3847/2041-8213/ace5a0},
  \href {https://ui.adsabs.harvard.edu/abs/2023ApJ...954L...4K} {954, L4}

\bibitem[\protect\citeauthoryear{{Kormendy} \& {Kennicutt}}{{Kormendy} \&
  {Kennicutt}}{2004}]{Kormendy2004}
{Kormendy} J.,  {Kennicutt} Robert~C. J.,  2004, \mn@doi [\araa]
  {10.1146/annurev.astro.42.053102.134024}, \href
  {https://ui.adsabs.harvard.edu/abs/2004ARA&A..42..603K} {42, 603}

\bibitem[\protect\citeauthoryear{{Lapiner}, {Dekel}  \& {Dubois}}{{Lapiner}
  et~al.}{2021}]{Lapiner2021}
{Lapiner} S.,  {Dekel} A.,   {Dubois} Y.,  2021, \mn@doi [\mnras]
  {10.1093/mnras/stab1205}, \href
  {https://ui.adsabs.harvard.edu/abs/2021MNRAS.505..172L} {505, 172}

\bibitem[\protect\citeauthoryear{{L{\"a}sker}, {Greene}, {Seth}, {van de Ven},
  {Braatz}, {Henkel}  \& {Lo}}{{L{\"a}sker} et~al.}{2016}]{Lasker2016}
{L{\"a}sker} R.,  {Greene} J.~E.,  {Seth} A.,  {van de Ven} G.,  {Braatz}
  J.~A.,  {Henkel} C.,   {Lo} K.~Y.,  2016, \mn@doi [\apj]
  {10.3847/0004-637X/825/1/3}, \href
  {https://ui.adsabs.harvard.edu/abs/2016ApJ...825....3L} {825, 3}

\bibitem[\protect\citeauthoryear{{Leitherer} et~al.,}{{Leitherer}
  et~al.}{1999}]{Leitherer1999}
{Leitherer} C.,  et~al., 1999, \mn@doi [\apjs] {10.1086/313233}, \href
  {https://ui.adsabs.harvard.edu/abs/1999ApJS..123....3L} {123, 3}

\bibitem[\protect\citeauthoryear{{Leitner} \& {Kravtsov}}{{Leitner} \&
  {Kravtsov}}{2011}]{Leitner2011}
{Leitner} S.~N.,  {Kravtsov} A.~V.,  2011, \mn@doi [\apj]
  {10.1088/0004-637X/734/1/48}, \href
  {https://ui.adsabs.harvard.edu/abs/2011ApJ...734...48L} {734, 48}

\bibitem[\protect\citeauthoryear{{Levine}, {Gnedin}, {Hamilton}  \&
  {Kravtsov}}{{Levine} et~al.}{2008}]{Levine2008}
{Levine} R.,  {Gnedin} N.~Y.,  {Hamilton} A. J.~S.,   {Kravtsov} A.~V.,  2008,
  \mn@doi [\apj] {10.1086/529064}, \href
  {https://ui.adsabs.harvard.edu/abs/2008ApJ...678..154L} {678, 154}

\bibitem[\protect\citeauthoryear{{Li}, {Bryan}, {Ruszkowski}, {Voit}, {O'Shea}
  \& {Donahue}}{{Li} et~al.}{2015}]{Li2015}
{Li} Y.,  {Bryan} G.~L.,  {Ruszkowski} M.,  {Voit} G.~M.,  {O'Shea} B.~W.,
  {Donahue} M.,  2015, \mn@doi [\apj] {10.1088/0004-637X/811/2/73}, \href
  {https://ui.adsabs.harvard.edu/abs/2015ApJ...811...73L} {811, 73}

\bibitem[\protect\citeauthoryear{{Looser} et~al.,}{{Looser}
  et~al.}{2024}]{Looser2024}
{Looser} T.~J.,  et~al., 2024, \mn@doi [\nat] {10.1038/s41586-024-07227-0},
  \href {https://ui.adsabs.harvard.edu/abs/2024Natur.629...53L} {629, 53}

\bibitem[\protect\citeauthoryear{{Looser} et~al.,}{{Looser}
  et~al.}{2025}]{Looser2025}
{Looser} T.~J.,  et~al., 2025, \mn@doi [\aap] {10.1051/0004-6361/202347102},
  \href {https://ui.adsabs.harvard.edu/abs/2025A&A...697A..88L} {697, A88}

\bibitem[\protect\citeauthoryear{{Ma}, {Hopkins}, {Feldmann}, {Torrey},
  {Faucher-Gigu{\`e}re}  \& {Kere{\v{s}}}}{{Ma} et~al.}{2017a}]{Ma2017a}
{Ma} X.,  {Hopkins} P.~F.,  {Feldmann} R.,  {Torrey} P.,  {Faucher-Gigu{\`e}re}
  C.-A.,   {Kere{\v{s}}} D.,  2017a, \mn@doi [\mnras] {10.1093/mnras/stx034},
  \href {https://ui.adsabs.harvard.edu/abs/2017MNRAS.466.4780M} {466, 4780}

\bibitem[\protect\citeauthoryear{{Ma}, {Hopkins}, {Wetzel}, {Kirby},
  {Angl{\'e}s-Alc{\'a}zar}, {Faucher-Gigu{\`e}re}, {Kere{\v{s}}}  \&
  {Quataert}}{{Ma} et~al.}{2017b}]{Ma2017b}
{Ma} X.,  {Hopkins} P.~F.,  {Wetzel} A.~R.,  {Kirby} E.~N.,
  {Angl{\'e}s-Alc{\'a}zar} D.,  {Faucher-Gigu{\`e}re} C.-A.,  {Kere{\v{s}}} D.,
    {Quataert} E.,  2017b, \mn@doi [\mnras] {10.1093/mnras/stx273}, \href
  {https://ui.adsabs.harvard.edu/abs/2017MNRAS.467.2430M} {467, 2430}

\bibitem[\protect\citeauthoryear{{Maiolino} et~al.,}{{Maiolino}
  et~al.}{2024}]{Maiolino2024}
{Maiolino} R.,  et~al., 2024, \mn@doi [\nat] {10.1038/s41586-024-07052-5},
  \href {https://ui.adsabs.harvard.edu/abs/2024Natur.627...59M} {627, 59}

\bibitem[\protect\citeauthoryear{{Martin}}{{Martin}}{2005}]{Martin2005}
{Martin} C.~L.,  2005, \mn@doi [\apj] {10.1086/427277}, \href
  {https://ui.adsabs.harvard.edu/abs/2005ApJ...621..227M} {621, 227}

\bibitem[\protect\citeauthoryear{{McAlpine}, {Harrison}, {Rosario},
  {Alexander}, {Ellison}, {Johansson}  \& {Patton}}{{McAlpine}
  et~al.}{2020}]{McAlpine2020}
{McAlpine} S.,  {Harrison} C.~M.,  {Rosario} D.~J.,  {Alexander} D.~M.,
  {Ellison} S.~L.,  {Johansson} P.~H.,   {Patton} D.~R.,  2020, \mn@doi
  [\mnras] {10.1093/mnras/staa1123}, \href
  {https://ui.adsabs.harvard.edu/abs/2020MNRAS.494.5713M} {494, 5713}

\bibitem[\protect\citeauthoryear{{Mercedes-Feliz} et~al.,}{{Mercedes-Feliz}
  et~al.}{2023}]{Mercedes-Feliz2023}
{Mercedes-Feliz} J.,  et~al., 2023, \mn@doi [\mnras] {10.1093/mnras/stad2079},
  \href {https://ui.adsabs.harvard.edu/abs/2023MNRAS.524.3446M} {524, 3446}

\bibitem[\protect\citeauthoryear{{Mercedes-Feliz} et~al.,}{{Mercedes-Feliz}
  et~al.}{2024}]{Mercedes-Feliz2024}
{Mercedes-Feliz} J.,  et~al., 2024, \mn@doi [\mnras] {10.1093/mnras/stae1021},
  \href {https://ui.adsabs.harvard.edu/abs/2024MNRAS.530.2795M} {530, 2795}

\bibitem[\protect\citeauthoryear{{Mintz} et~al.,}{{Mintz}
  et~al.}{2025}]{Mintz2025}
{Mintz} A.,  et~al., 2025, \mn@doi [arXiv e-prints]
  {10.48550/arXiv.2506.16510}, \href
  {https://ui.adsabs.harvard.edu/abs/2025arXiv250616510M} {p. arXiv:2506.16510}

\bibitem[\protect\citeauthoryear{{Mitchell}, {Schaye}, {Bower}  \&
  {Crain}}{{Mitchell} et~al.}{2020}]{Mitchell2020}
{Mitchell} P.~D.,  {Schaye} J.,  {Bower} R.~G.,   {Crain} R.~A.,  2020, \mn@doi
  [\mnras] {10.1093/mnras/staa938}, \href
  {https://ui.adsabs.harvard.edu/abs/2020MNRAS.494.3971M} {494, 3971}

\bibitem[\protect\citeauthoryear{{Muratov}, {Kere{\v{s}}},
  {Faucher-Gigu{\`e}re}, {Hopkins}, {Quataert}  \& {Murray}}{{Muratov}
  et~al.}{2015}]{Muratov2015}
{Muratov} A.~L.,  {Kere{\v{s}}} D.,  {Faucher-Gigu{\`e}re} C.-A.,  {Hopkins}
  P.~F.,  {Quataert} E.,   {Murray} N.,  2015, \mn@doi [\mnras]
  {10.1093/mnras/stv2126}, \href
  {https://ui.adsabs.harvard.edu/abs/2015MNRAS.454.2691M} {454, 2691}

\bibitem[\protect\citeauthoryear{{Muratov} et~al.,}{{Muratov}
  et~al.}{2017}]{Muratov2017}
{Muratov} A.~L.,  et~al., 2017, \mn@doi [\mnras] {10.1093/mnras/stx667}, \href
  {https://ui.adsabs.harvard.edu/abs/2017MNRAS.468.4170M} {468, 4170}

\bibitem[\protect\citeauthoryear{{Nelson} et~al.,}{{Nelson}
  et~al.}{2019}]{Nelson2019}
{Nelson} D.,  et~al., 2019, \mn@doi [\mnras] {10.1093/mnras/stz2306}, \href
  {https://ui.adsabs.harvard.edu/abs/2019MNRAS.490.3234N} {490, 3234}

\bibitem[\protect\citeauthoryear{{Nguyen} et~al.,}{{Nguyen}
  et~al.}{2019}]{Nguyen2019}
{Nguyen} D.~D.,  et~al., 2019, \mn@doi [\apj] {10.3847/1538-4357/aafe7a}, \href
  {https://ui.adsabs.harvard.edu/abs/2019ApJ...872..104N} {872, 104}

\bibitem[\protect\citeauthoryear{{Novak}, {Ostriker}  \& {Ciotti}}{{Novak}
  et~al.}{2011}]{Novak2011}
{Novak} G.~S.,  {Ostriker} J.~P.,   {Ciotti} L.,  2011, \mn@doi [\apj]
  {10.1088/0004-637X/737/1/26}, \href
  {https://ui.adsabs.harvard.edu/abs/2011ApJ...737...26N} {737, 26}

\bibitem[\protect\citeauthoryear{{Nyland} et~al.,}{{Nyland}
  et~al.}{2020}]{Nyland2020}
{Nyland} K.,  et~al., 2020, \mn@doi [\apj] {10.3847/1538-4357/abc341}, \href
  {https://ui.adsabs.harvard.edu/abs/2020ApJ...905...74N} {905, 74}

\bibitem[\protect\citeauthoryear{{Oppenheimer}, {Dav{\'e}}, {Kere{\v{s}}},
  {Fardal}, {Katz}, {Kollmeier}  \& {Weinberg}}{{Oppenheimer}
  et~al.}{2010}]{Oppenheimer2010}
{Oppenheimer} B.~D.,  {Dav{\'e}} R.,  {Kere{\v{s}}} D.,  {Fardal} M.,  {Katz}
  N.,  {Kollmeier} J.~A.,   {Weinberg} D.~H.,  2010, \mn@doi [\mnras]
  {10.1111/j.1365-2966.2010.16872.x}, \href
  {https://ui.adsabs.harvard.edu/abs/2010MNRAS.406.2325O} {406, 2325}

\bibitem[\protect\citeauthoryear{{Oser}, {Ostriker}, {Naab}, {Johansson}  \&
  {Burkert}}{{Oser} et~al.}{2010}]{Oser2010}
{Oser} L.,  {Ostriker} J.~P.,  {Naab} T.,  {Johansson} P.~H.,   {Burkert} A.,
  2010, \mn@doi [\apj] {10.1088/0004-637X/725/2/2312}, \href
  {https://ui.adsabs.harvard.edu/abs/2010ApJ...725.2312O} {725, 2312}

\bibitem[\protect\citeauthoryear{{Padovani} et~al.,}{{Padovani}
  et~al.}{2017}]{Padovani2017}
{Padovani} P.,  et~al., 2017, \mn@doi [\aapr] {10.1007/s00159-017-0102-9},
  \href {https://ui.adsabs.harvard.edu/abs/2017A&ARv..25....2P} {25, 2}

\bibitem[\protect\citeauthoryear{{Pandya} et~al.,}{{Pandya}
  et~al.}{2021}]{Pandya2021}
{Pandya} V.,  et~al., 2021, \mn@doi [\mnras] {10.1093/mnras/stab2714}, \href
  {https://ui.adsabs.harvard.edu/abs/2021MNRAS.508.2979P} {508, 2979}

\bibitem[\protect\citeauthoryear{{Parsotan}, {Cochrane}, {Hayward},
  {Angl{\'e}s-Alc{\'a}zar}, {Feldmann}, {Faucher-Gigu{\`e}re}, {Wellons}  \&
  {Hopkins}}{{Parsotan} et~al.}{2021}]{Parsotan2021}
{Parsotan} T.,  {Cochrane} R.~K.,  {Hayward} C.~C.,  {Angl{\'e}s-Alc{\'a}zar}
  D.,  {Feldmann} R.,  {Faucher-Gigu{\`e}re} C.~A.,  {Wellons} S.,   {Hopkins}
  P.~F.,  2021, \mn@doi [\mnras] {10.1093/mnras/staa3765}, \href
  {https://ui.adsabs.harvard.edu/abs/2021MNRAS.501.1591P} {501, 1591}

\bibitem[\protect\citeauthoryear{{Pierce} et~al.,}{{Pierce}
  et~al.}{2007}]{Pierce2007}
{Pierce} C.~M.,  et~al., 2007, \mn@doi [\apjl] {10.1086/517922}, \href
  {https://ui.adsabs.harvard.edu/abs/2007ApJ...660L..19P} {660, L19}

\bibitem[\protect\citeauthoryear{{Pillepich} et~al.,}{{Pillepich}
  et~al.}{2018}]{Pillepich2018}
{Pillepich} A.,  et~al., 2018, \mnras, 473, 4077

\bibitem[\protect\citeauthoryear{{Pontzen}, {Tremmel}, {Roth}, {Peiris},
  {Saintonge}, {Volonteri}, {Quinn}  \& {Governato}}{{Pontzen}
  et~al.}{2017}]{Pontzen2017}
{Pontzen} A.,  {Tremmel} M.,  {Roth} N.,  {Peiris} H.~V.,  {Saintonge} A.,
  {Volonteri} M.,  {Quinn} T.,   {Governato} F.,  2017, \mn@doi [\mnras]
  {10.1093/mnras/stw2627}, \href
  {https://ui.adsabs.harvard.edu/abs/2017MNRAS.465..547P} {465, 547}

\bibitem[\protect\citeauthoryear{{Reines} \& {Volonteri}}{{Reines} \&
  {Volonteri}}{2015}]{Reines2015}
{Reines} A.~E.,  {Volonteri} M.,  2015, \mn@doi [\apj]
  {10.1088/0004-637X/813/2/82}, \href
  {https://ui.adsabs.harvard.edu/abs/2015ApJ...813...82R} {813, 82}

\bibitem[\protect\citeauthoryear{{Ricarte}, {Tremmel}, {Natarajan}  \&
  {Quinn}}{{Ricarte} et~al.}{2019}]{Ricarte2019}
{Ricarte} A.,  {Tremmel} M.,  {Natarajan} P.,   {Quinn} T.,  2019, \mn@doi
  [\mnras] {10.1093/mnras/stz2161}, \href
  {https://ui.adsabs.harvard.edu/abs/2019MNRAS.489..802R} {489, 802}

\bibitem[\protect\citeauthoryear{{Riffel} et~al.,}{{Riffel}
  et~al.}{2023}]{Riffel2023}
{Riffel} R.,  et~al., 2023, \mn@doi [\mnras] {10.1093/mnras/stad2234}, \href
  {https://ui.adsabs.harvard.edu/abs/2023MNRAS.524.5640R} {524, 5640}

\bibitem[\protect\citeauthoryear{{Roberts-Borsani}, {Saintonge}, {Masters}  \&
  {Stark}}{{Roberts-Borsani} et~al.}{2020}]{Roberts-Borsani2020}
{Roberts-Borsani} G.~W.,  {Saintonge} A.,  {Masters} K.~L.,   {Stark} D.~V.,
  2020, \mn@doi [\mnras] {10.1093/mnras/staa464}, \href
  {https://ui.adsabs.harvard.edu/abs/2020MNRAS.493.3081R} {493, 3081}

\bibitem[\protect\citeauthoryear{{Rodr{\'\i}guez-Puebla}, {Primack}, {Behroozi}
   \& {Faber}}{{Rodr{\'\i}guez-Puebla} et~al.}{2016}]{Rodriguez-Puebla2016}
{Rodr{\'\i}guez-Puebla} A.,  {Primack} J.~R.,  {Behroozi} P.,   {Faber} S.~M.,
  2016, \mn@doi [\mnras] {10.1093/mnras/stv2513}, \href
  {https://ui.adsabs.harvard.edu/abs/2016MNRAS.455.2592R} {455, 2592}

\bibitem[\protect\citeauthoryear{{Rubin}, {Prochaska}, {Koo}, {Phillips},
  {Martin}  \& {Winstrom}}{{Rubin} et~al.}{2014}]{Rubin2014}
{Rubin} K. H.~R.,  {Prochaska} J.~X.,  {Koo} D.~C.,  {Phillips} A.~C.,
  {Martin} C.~L.,   {Winstrom} L.~O.,  2014, \mn@doi [\apj]
  {10.1088/0004-637X/794/2/156}, \href
  {https://ui.adsabs.harvard.edu/abs/2014ApJ...794..156R} {794, 156}

\bibitem[\protect\citeauthoryear{{Sahu}, {Graham}  \& {Davis}}{{Sahu}
  et~al.}{2019}]{Sahu2019}
{Sahu} N.,  {Graham} A.~W.,   {Davis} B.~L.,  2019, \mn@doi [\apj]
  {10.3847/1538-4357/ab0f32}, \href
  {https://ui.adsabs.harvard.edu/abs/2019ApJ...876..155S} {876, 155}

\bibitem[\protect\citeauthoryear{{Salvador-Rusi{\~n}ol}, {Beasley}, {Vazdekis}
  \& {Barbera}}{{Salvador-Rusi{\~n}ol} et~al.}{2021}]{Salvador-Rusinol2021}
{Salvador-Rusi{\~n}ol} N.,  {Beasley} M.~A.,  {Vazdekis} A.,   {Barbera} F.~L.,
   2021, \mn@doi [\mnras] {10.1093/mnras/staa3419}, \href
  {https://ui.adsabs.harvard.edu/abs/2021MNRAS.500.3368S} {500, 3368}

\bibitem[\protect\citeauthoryear{{Sanders} et~al.,}{{Sanders}
  et~al.}{2016}]{Sanders2016}
{Sanders} R.~L.,  et~al., 2016, \mn@doi [\apj] {10.3847/0004-637X/816/1/23},
  \href {https://ui.adsabs.harvard.edu/abs/2016ApJ...816...23S} {816, 23}

\bibitem[\protect\citeauthoryear{{Savorgnan}, {Graham}, {Marconi}  \&
  {Sani}}{{Savorgnan} et~al.}{2016}]{Savorgnan2016}
{Savorgnan} G. A.~D.,  {Graham} A.~W.,  {Marconi} A.,   {Sani} E.,  2016,
  \mn@doi [\apj] {10.3847/0004-637X/817/1/21}, \href
  {https://ui.adsabs.harvard.edu/abs/2016ApJ...817...21S} {817, 21}

\bibitem[\protect\citeauthoryear{{Saxena} et~al.,}{{Saxena}
  et~al.}{2024}]{Saxena2024}
{Saxena} A.,  et~al., 2024, \mn@doi [\mnras] {10.1093/mnras/stae1406}, \href
  {https://ui.adsabs.harvard.edu/abs/2024MNRAS.531.4391S} {531, 4391}

\bibitem[\protect\citeauthoryear{{Schawinski}, {Koss}, {Berney}  \&
  {Sartori}}{{Schawinski} et~al.}{2015}]{Schawinski2015}
{Schawinski} K.,  {Koss} M.,  {Berney} S.,   {Sartori} L.~F.,  2015, \mn@doi
  [\mnras] {10.1093/mnras/stv1136}, \href
  {https://ui.adsabs.harvard.edu/abs/2015MNRAS.451.2517S} {451, 2517}

\bibitem[\protect\citeauthoryear{{Schinnerer} et~al.,}{{Schinnerer}
  et~al.}{2023}]{Schinnerer2023}
{Schinnerer} E.,  et~al., 2023, \mn@doi [\apjl] {10.3847/2041-8213/acac9e},
  \href {https://ui.adsabs.harvard.edu/abs/2023ApJ...944L..15S} {944, L15}

\bibitem[\protect\citeauthoryear{{Schneider}, {Ostriker}, {Robertson}  \&
  {Thompson}}{{Schneider} et~al.}{2020}]{Schneider2020}
{Schneider} E.~E.,  {Ostriker} E.~C.,  {Robertson} B.~E.,   {Thompson} T.~A.,
  2020, \mn@doi [\apj] {10.3847/1538-4357/ab8ae8}, \href
  {https://ui.adsabs.harvard.edu/abs/2020ApJ...895...43S} {895, 43}

\bibitem[\protect\citeauthoryear{{Schutte}, {Reines}  \& {Greene}}{{Schutte}
  et~al.}{2019}]{Schutte2019}
{Schutte} Z.,  {Reines} A.~E.,   {Greene} J.~E.,  2019, \mn@doi [\apj]
  {10.3847/1538-4357/ab35dd}, \href
  {https://ui.adsabs.harvard.edu/abs/2019ApJ...887..245S} {887, 245}

\bibitem[\protect\citeauthoryear{{Segers}, {Crain}, {Schaye}, {Bower},
  {Furlong}, {Schaller}  \& {Theuns}}{{Segers} et~al.}{2016}]{Segers2016}
{Segers} M.~C.,  {Crain} R.~A.,  {Schaye} J.,  {Bower} R.~G.,  {Furlong} M.,
  {Schaller} M.,   {Theuns} T.,  2016, \mn@doi [\mnras]
  {10.1093/mnras/stv2562}, \href
  {https://ui.adsabs.harvard.edu/abs/2016MNRAS.456.1235S} {456, 1235}

\bibitem[\protect\citeauthoryear{{Sharma} et~al.,}{{Sharma}
  et~al.}{2024}]{Sharma2024}
{Sharma} R.~S.,  et~al., 2024, \mn@doi [\mnras] {10.1093/mnras/stad3836}, \href
  {https://ui.adsabs.harvard.edu/abs/2024MNRAS.527.9461S} {527, 9461}

\bibitem[\protect\citeauthoryear{{Shen}}{{Shen}}{2021}]{Shen2021}
{Shen} Y.,  2021, \mn@doi [\apj] {10.3847/1538-4357/ac1ce4}, \href
  {https://ui.adsabs.harvard.edu/abs/2021ApJ...921...70S} {921, 70}

\bibitem[\protect\citeauthoryear{{Shipp} et~al.,}{{Shipp}
  et~al.}{2023}]{Shipp2023}
{Shipp} N.,  et~al., 2023, \mn@doi [\apj] {10.3847/1538-4357/acc582}, \href
  {https://ui.adsabs.harvard.edu/abs/2023ApJ...949...44S} {949, 44}

\bibitem[\protect\citeauthoryear{{Simmonds} et~al.,}{{Simmonds}
  et~al.}{2025}]{Simmonds2025}
{Simmonds} C.,  et~al., 2025, \mn@doi [arXiv e-prints]
  {10.48550/arXiv.2508.04410}, \href
  {https://ui.adsabs.harvard.edu/abs/2025arXiv250804410S} {p. arXiv:2508.04410}

\bibitem[\protect\citeauthoryear{{Smith}, {Sijacki}  \& {Shen}}{{Smith}
  et~al.}{2018}]{Smith2018}
{Smith} M.~C.,  {Sijacki} D.,   {Shen} S.,  2018, \mn@doi [\mnras]
  {10.1093/mnras/sty994}, \href
  {https://ui.adsabs.harvard.edu/abs/2018MNRAS.478..302S} {478, 302}

\bibitem[\protect\citeauthoryear{{Sparre}, {Hayward}, {Feldmann},
  {Faucher-Gigu{\`e}re}, {Muratov}, {Kere{\v{s}}}  \& {Hopkins}}{{Sparre}
  et~al.}{2017}]{Sparre2017}
{Sparre} M.,  {Hayward} C.~C.,  {Feldmann} R.,  {Faucher-Gigu{\`e}re} C.-A.,
  {Muratov} A.~L.,  {Kere{\v{s}}} D.,   {Hopkins} P.~F.,  2017, \mn@doi
  [\mnras] {10.1093/mnras/stw3011}, \href
  {https://ui.adsabs.harvard.edu/abs/2017MNRAS.466...88S} {466, 88}

\bibitem[\protect\citeauthoryear{{Springel}, {Di Matteo}  \&
  {Hernquist}}{{Springel} et~al.}{2005}]{Springel2005}
{Springel} V.,  {Di Matteo} T.,   {Hernquist} L.,  2005, \mn@doi [\apjl]
  {10.1086/428772}, \href
  {https://ui.adsabs.harvard.edu/abs/2005ApJ...620L..79S} {620, L79}

\bibitem[\protect\citeauthoryear{{Steinwandel}, {Kim}, {Bryan}, {Ostriker},
  {Somerville}  \& {Fielding}}{{Steinwandel} et~al.}{2024}]{Steinwandel2024}
{Steinwandel} U.~P.,  {Kim} C.-G.,  {Bryan} G.~L.,  {Ostriker} E.~C.,
  {Somerville} R.~S.,   {Fielding} D.~B.,  2024, \mn@doi [\apj]
  {10.3847/1538-4357/ad09e1}, \href
  {https://ui.adsabs.harvard.edu/abs/2024ApJ...960..100S} {960, 100}

\bibitem[\protect\citeauthoryear{{Stern} et~al.,}{{Stern}
  et~al.}{2021}]{Stern2021}
{Stern} J.,  et~al., 2021, \mn@doi [\apj] {10.3847/1538-4357/abd776}, \href
  {https://ui.adsabs.harvard.edu/abs/2021ApJ...911...88S} {911, 88}

\bibitem[\protect\citeauthoryear{{Tacchella}, {Dekel}, {Carollo}, {Ceverino},
  {DeGraf}, {Lapiner}, {Mandelker}  \& {Primack Joel}}{{Tacchella}
  et~al.}{2016}]{Tacchella2016}
{Tacchella} S.,  {Dekel} A.,  {Carollo} C.~M.,  {Ceverino} D.,  {DeGraf} C.,
  {Lapiner} S.,  {Mandelker} N.,   {Primack Joel} R.,  2016, \mn@doi [\mnras]
  {10.1093/mnras/stw131}, \href
  {https://ui.adsabs.harvard.edu/abs/2016MNRAS.457.2790T} {457, 2790}

\bibitem[\protect\citeauthoryear{{Tacchella}, {Forbes}  \&
  {Caplar}}{{Tacchella} et~al.}{2020}]{Tacchella2020}
{Tacchella} S.,  {Forbes} J.~C.,   {Caplar} N.,  2020, \mn@doi [\mnras]
  {10.1093/mnras/staa1838}, \href
  {https://ui.adsabs.harvard.edu/abs/2020MNRAS.497..698T} {497, 698}

\bibitem[\protect\citeauthoryear{{Talbot}, {Sijacki}  \& {Bourne}}{{Talbot}
  et~al.}{2022}]{Talbot2022}
{Talbot} R.~Y.,  {Sijacki} D.,   {Bourne} M.~A.,  2022, \mn@doi [\mnras]
  {10.1093/mnras/stac1566}, \href
  {https://ui.adsabs.harvard.edu/abs/2022MNRAS.514.4535T} {514, 4535}

\bibitem[\protect\citeauthoryear{{Tillman}, {Wellons}, {Faucher-Gigu{\`e}re},
  {Kelley}  \& {Angl{\'e}s-Alc{\'a}zar}}{{Tillman} et~al.}{2022}]{Tillman2022}
{Tillman} M.~T.,  {Wellons} S.,  {Faucher-Gigu{\`e}re} C.-A.,  {Kelley} L.~Z.,
   {Angl{\'e}s-Alc{\'a}zar} D.,  2022, \mn@doi [\mnras]
  {10.1093/mnras/stac398}, \href
  {https://ui.adsabs.harvard.edu/abs/2022MNRAS.511.5756T} {511, 5756}

\bibitem[\protect\citeauthoryear{{Tollet}, {Cattaneo}, {Macci{\`o}}, {Dutton}
  \& {Kang}}{{Tollet} et~al.}{2019}]{Tollet2019}
{Tollet} {\'E}.,  {Cattaneo} A.,  {Macci{\`o}} A.~V.,  {Dutton} A.~A.,   {Kang}
  X.,  2019, \mn@doi [\mnras] {10.1093/mnras/stz545}, \href
  {https://ui.adsabs.harvard.edu/abs/2019MNRAS.485.2511T} {485, 2511}

\bibitem[\protect\citeauthoryear{{Treister}, {Schawinski}, {Urry}  \&
  {Simmons}}{{Treister} et~al.}{2012}]{Treister2012}
{Treister} E.,  {Schawinski} K.,  {Urry} C.~M.,   {Simmons} B.~D.,  2012,
  \mn@doi [\apjl] {10.1088/2041-8205/758/2/L39}, \href
  {https://ui.adsabs.harvard.edu/abs/2012ApJ...758L..39T} {758, L39}

\bibitem[\protect\citeauthoryear{{Tremmel} et~al.,}{{Tremmel}
  et~al.}{2019}]{Tremmel2019}
{Tremmel} M.,  et~al., 2019, \mn@doi [\mnras] {10.1093/mnras/sty3336}, \href
  {https://ui.adsabs.harvard.edu/abs/2019MNRAS.483.3336T} {483, 3336}

\bibitem[\protect\citeauthoryear{{{\"U}bler} et~al.,}{{{\"U}bler}
  et~al.}{2023}]{Ubler2023}
{{\"U}bler} H.,  et~al., 2023, \mn@doi [\aap] {10.1051/0004-6361/202346137},
  \href {https://ui.adsabs.harvard.edu/abs/2023A&A...677A.145U} {677, A145}

\bibitem[\protect\citeauthoryear{{Veilleux}, {Kim}  \& {Sanders}}{{Veilleux}
  et~al.}{2002}]{Veilleux2002}
{Veilleux} S.,  {Kim} D.~C.,   {Sanders} D.~B.,  2002, \mn@doi [\apjs]
  {10.1086/343844}, \href
  {https://ui.adsabs.harvard.edu/abs/2002ApJS..143..315V} {143, 315}

\bibitem[\protect\citeauthoryear{{Villforth}}{{Villforth}}{2023}]{Villforth2023}
{Villforth} C.,  2023, \mn@doi [The Open Journal of Astrophysics]
  {10.21105/astro.2309.03276}, \href
  {https://ui.adsabs.harvard.edu/abs/2023OJAp....6E..34V} {6, 34}

\bibitem[\protect\citeauthoryear{{Villforth} et~al.,}{{Villforth}
  et~al.}{2014}]{Villforth2014}
{Villforth} C.,  et~al., 2014, \mn@doi [\mnras] {10.1093/mnras/stu173}, \href
  {https://ui.adsabs.harvard.edu/abs/2014MNRAS.439.3342V} {439, 3342}

\bibitem[\protect\citeauthoryear{{Villforth}, {Herbst}, {Hamann}, {Hamilton},
  {Bertemes}, {Efthymiadou}  \& {Hewlett}}{{Villforth}
  et~al.}{2019}]{Villforth2019}
{Villforth} C.,  {Herbst} H.,  {Hamann} F.,  {Hamilton} T.,  {Bertemes} C.,
  {Efthymiadou} A.,   {Hewlett} T.,  2019, \mn@doi [\mnras]
  {10.1093/mnras/sty3271}, \href
  {https://ui.adsabs.harvard.edu/abs/2019MNRAS.483.2441V} {483, 2441}

\bibitem[\protect\citeauthoryear{{Voit}, {Donahue}, {Bryan}  \&
  {McDonald}}{{Voit} et~al.}{2015}]{Voit2015}
{Voit} G.~M.,  {Donahue} M.,  {Bryan} G.~L.,   {McDonald} M.,  2015, \mn@doi
  [\nat] {10.1038/nature14167}, \href
  {https://ui.adsabs.harvard.edu/abs/2015Natur.519..203V} {519, 203}

\bibitem[\protect\citeauthoryear{{Weisz} et~al.,}{{Weisz}
  et~al.}{2012}]{Weisz2012}
{Weisz} D.~R.,  et~al., 2012, \mn@doi [\apj] {10.1088/0004-637X/744/1/44},
  \href {https://ui.adsabs.harvard.edu/abs/2012ApJ...744...44W} {744, 44}

\bibitem[\protect\citeauthoryear{{Wellons}, {Faucher-Gigu{\`e}re},
  {Angl{\'e}s-Alc{\'a}zar}, {Hayward}, {Feldmann}, {Hopkins}  \&
  {Kere{\v{s}}}}{{Wellons} et~al.}{2020}]{Wellons2020}
{Wellons} S.,  {Faucher-Gigu{\`e}re} C.-A.,  {Angl{\'e}s-Alc{\'a}zar} D.,
  {Hayward} C.~C.,  {Feldmann} R.,  {Hopkins} P.~F.,   {Kere{\v{s}}} D.,  2020,
  \mn@doi [\mnras] {10.1093/mnras/staa2229}, \href
  {https://ui.adsabs.harvard.edu/abs/2020MNRAS.497.4051W} {497, 4051}

\bibitem[\protect\citeauthoryear{{Wellons} et~al.,}{{Wellons}
  et~al.}{2023}]{Wellons2023}
{Wellons} S.,  et~al., 2023, \mn@doi [\mnras] {10.1093/mnras/stad511}, \href
  {https://ui.adsabs.harvard.edu/abs/2023MNRAS.520.5394W} {520, 5394}

\bibitem[\protect\citeauthoryear{{Wetzel} et~al.,}{{Wetzel}
  et~al.}{2023}]{Wetzel2023}
{Wetzel} A.,  et~al., 2023, \mn@doi [\apjs] {10.3847/1538-4365/acb99a}, \href
  {https://ui.adsabs.harvard.edu/abs/2023ApJS..265...44W} {265, 44}

\bibitem[\protect\citeauthoryear{{Wetzel} et~al.,}{{Wetzel}
  et~al.}{2025}]{Wetzel2025}
{Wetzel} A.,  et~al., 2025, \mn@doi [arXiv e-prints]
  {10.48550/arXiv.2508.06608}, \href
  {https://ui.adsabs.harvard.edu/abs/2025arXiv250806608W} {p. arXiv:2508.06608}

\bibitem[\protect\citeauthoryear{{Yu} et~al.,}{{Yu} et~al.}{2021}]{Yu2021}
{Yu} S.,  et~al., 2021, \mn@doi [\mnras] {10.1093/mnras/stab1339}, \href
  {https://ui.adsabs.harvard.edu/abs/2021MNRAS.505..889Y} {505, 889}

\bibitem[\protect\citeauthoryear{{{\c{C}}atmabacak}, {Feldmann},
  {Angl{\'e}s-Alc{\'a}zar}, {Faucher-Gigu{\`e}re}, {Hopkins}  \&
  {Kere{\v{s}}}}{{{\c{C}}atmabacak} et~al.}{2022}]{Catmabacak2022}
{{\c{C}}atmabacak} O.,  {Feldmann} R.,  {Angl{\'e}s-Alc{\'a}zar} D.,
  {Faucher-Gigu{\`e}re} C.-A.,  {Hopkins} P.~F.,   {Kere{\v{s}}} D.,  2022,
  \mn@doi [\mnras] {10.1093/mnras/stac040}, \href
  {https://ui.adsabs.harvard.edu/abs/2022MNRAS.511..506C} {511, 506}

\bibitem[\protect\citeauthoryear{{de Graaff} et~al.,}{{de Graaff}
  et~al.}{2024}]{deGraaff2024}
{de Graaff} A.,  et~al., 2024, \mn@doi [\aap] {10.1051/0004-6361/202347755},
  \href {https://ui.adsabs.harvard.edu/abs/2024A&A...684A..87D} {684, A87}

\bibitem[\protect\citeauthoryear{{van de Voort}}{{van de
  Voort}}{2017}]{vandeVoort2017}
{van de Voort} F.,  2017, in {Fox} A.,  {Dav{\'e}} R.,  eds,  Astrophysics and
  Space Science Library Vol. 430, Gas Accretion onto Galaxies. p.~301
  (\mn@eprint {arXiv} {1612.00591}), \mn@doi{10.1007/978-3-319-52512-9_13}

\makeatother
\end{thebibliography}






\bsp	
\label{lastpage}
\end{document}